\documentclass[journal=acsearthspacechem,manuscript=article]{achemso}
\SectionNumbersOn

\UseRawInputEncoding 

\usepackage{ifthen}
\usepackage[version=3]{mhchem} 
\usepackage{multirow} 
\usepackage{xcolor}

\newboolean{showhighlights}
\setboolean{showhighlights}{false} 

\definecolor{zblue}{RGB}{14, 100, 190} 

\ifthenelse{\boolean{showhighlights}}%
  {\newcommand{\hg}[1]{\textcolor{zblue}{#1}}}%
  {\newcommand{\hg}[1]{#1}}

\author{Gabriela Silva-Vera}
\affiliation{Departamento de Físico-Química, Facultad de Ciencias Químicas, Universidad de Concepción,4070386 Concepción, Chile}
\author{Giulia M. Bovolenta}
\affiliation{Departamento de Físico-Química, Facultad de Ciencias Químicas, Universidad de Concepción,4070386 Concepción, Chile}
\altaffiliation{Atomistic Simulations, Italian Institute of Technology, 16152 Genova, Italy}
\author{Namrata Rani}
\affiliation{Departamento de Físico-Química, Facultad de Ciencias Químicas, Universidad de Concepción, 4070386 Concepción, Chile}
\author{Sebastian Vera}
\affiliation{Departamento de Físico-Química, Facultad de Ciencias Químicas, Universidad de Concepción,4070386 Concepción, Chile}
\author{Stefan Vogt-Geisse}
\affiliation{Departamento de Físico-Química, Facultad de Ciencias Químicas, Universidad de Concepción, 4070386 Concepción, Chile}
\email{stvogtgeisse@qcmmlab.com}

\title{Pathways to interstellar amides via carbamoyl (\ce{NH2CO}) isomers by radical-neutral reactions on 
 icy-grains}
 

\begin{document}

\begin{tocentry}

\includegraphics[width=\textwidth]{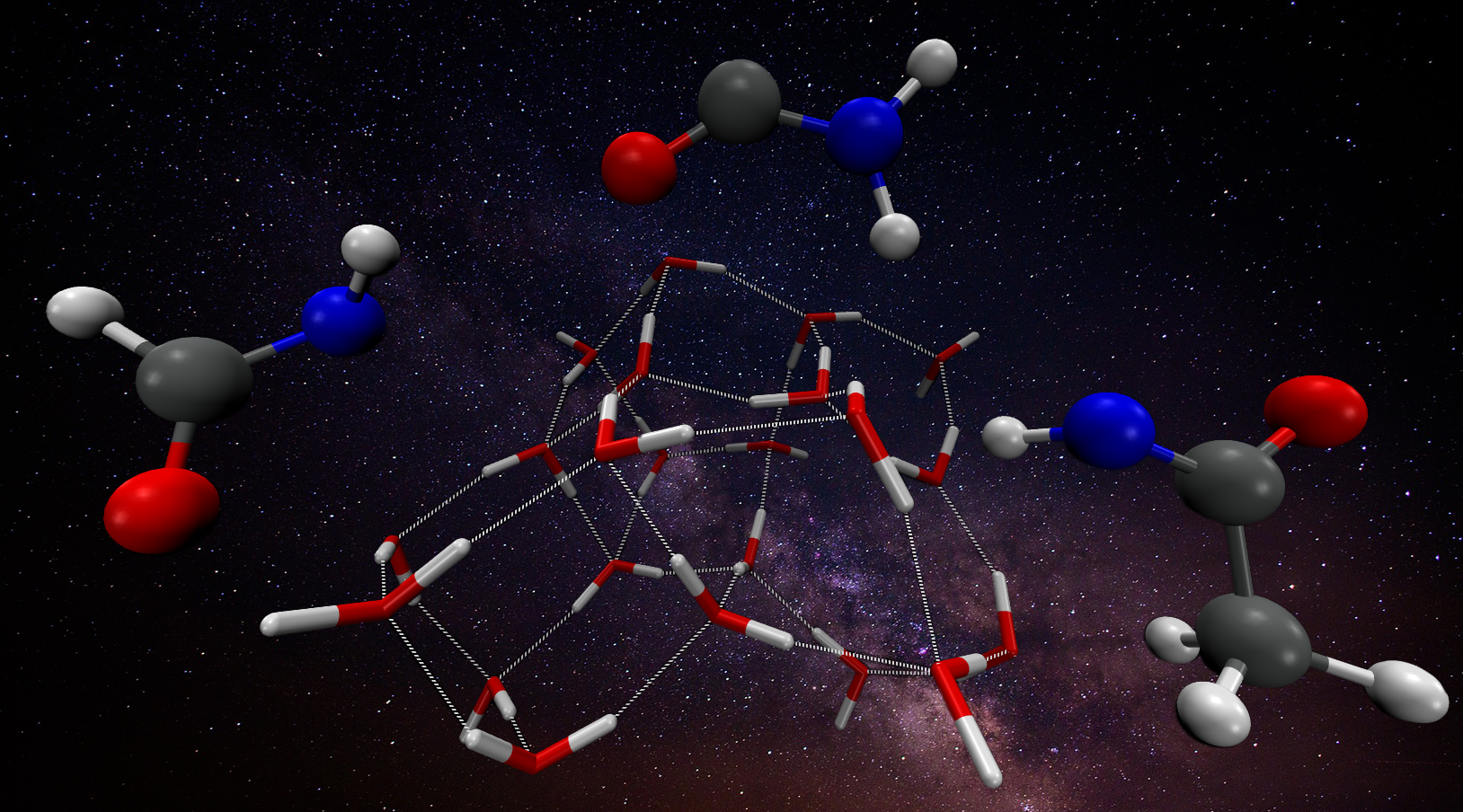}

\end{tocentry}

\clearpage
\begin{abstract}
Explaining the formation pathways of amides on ice-grain mantels
is crucial to the understanding of prebiotic chemistry in the 
interstellar medium. In this computational study, we explore different 
radical-neutral formation pathways for  some  of the observed amides 
(formamide, acetamide, urea and N-methylformamide) via intermediate 
carbamoyl  (\ce{NH2CO}) radical precursor and its isomers. We assess the 
relative energy of four \ce{NH2CO} isomers in the gas phase and evaluate 
their binding energy on small water clusters to discern the affinity that 
the isomers present to a ice model. We consider three possible reaction pathways 
for the formation of the carbamoyl radicals, namely the
\ce{OH + HCN}, \ce{CN + H2O} and \ce{NH2 + CO} reaction channels.  
We computed the binding energy distribution for the \ce{HCN} and \ce{CH3CN} precursors 
on a ice model consisting of a set-of-clusters  of 22-water molecules each,  
to serve as a starting point to the reactivity study on the ice surface. The 
computations revealed that the lowest barrier to the formation of a \ce{NH2CO} 
isomer corresponds to the \ce{NH2 + CO}  reaction (12.6 kJ $\mathrm{mol^{-1}}$). The reaction 
pathway \ce{OH + HCN} results in the exothermic  formation of the N-radical form 
of the carbamoyl  \ce{HN(C=O)H} with a reaction  barrier of 26.7 kJ $\mathrm{mol^{-1}}$. 
We found that the \ce{CN + H2O} reaction displays  a high energy barrier of 70.6 kJ $\mathrm{mol^{-1}}$. 
Finally, we also probed the direct formation of acetamide radical precursor  via the 
\ce{OH + CH3CN} reaction,  and found that the most probable outcome on interstellar ices 
is  the H-abstraction  reaction to yield \ce{CH2CN} and \ce{H2O}. Based on those result,  
we believe  that  including  alternative  reactions pathways,  leading to the formation of
amides via the N-radical form of carbamoyl, would provide an improvement in the prediction 
of the amides abundances in astrochemical models, especially regarding chemistry of star-forming 
regions.   
\end{abstract}

\clearpage

\section{Introduction}\label{sec:sec1}
\hg{Despite its overreaching emptiness, the interstellar medium (ISM) harbors gas,
dust, and organic compounds. This space includes molecular clouds, which are colder, denser areas with 
temperatures ranging from 10 to 30 K and particle densities between 10$^4$ and 10$^8$ particles per 
cubic centimeter.  Under these frigid conditions, gas-phase molecules adhere to dust grains - composites 
of non-volatile materials like carbon-based substances, oxides, and silicates - forming an icy mantle. 
Such ice mantles are rich in water and also contain a mix of simple molecules, including  
\ce{CO2}, \ce{CO}, \ce{CH4} and \ce{NH3}\cite{mcclure_ice_2023}. The formation of interstellar complex organic molecules 
i(iCOMs) is believed to take place on the surface of the ice mantles.} 
\hg{Among the iCOMs, amides are prebiotic compounds characterized by a carbonyl group bonded 
to a nitrogen atom ($\mathrm{R-C(=O)-N-R'}$). This structural characteristic, known as a peptide bond, 
is fundamental for the formation of 
life's building blocks like amino acids\cite{portugal_radiolysis_2014}}.  The majority of the 
observations of interstellar amides have been made towards star-forming regions\cite{lopez-sepulcre_interstellar_2019,colzi_guapos_2021,ligterink_family_2020,ligterink_prebiotic_2022}.   
Recently, the results of a sensitive spectral survey\cite{zeng_amides_2023} carried out towards the quiescent molecular cloud G+0.693-0.027 revealed a rich amide inventory. 
They found evidence of the presence of the simplest amides,
such as formamide (\ce{NH2CHO}), acetamide (\ce{CH3C(O)NH2}), N-methylformamide (\ce{CH3NHCHO}) and urea (\ce{NH2C(O)NH2}), while the quest for more complex amides is still ongoing in that as well as in other interstellar regions\cite{colzi_guapos_2021, ligterink_prebiotic_2022-2,sanz-novo_toward_2022}.
Another notable
result from the latest survey\cite{zeng_amides_2023}, was that the relative abundances of amides, when compared to formamide, 
typically exhibit the least difference across the different source environments, with variations 
frequently not exceeding a factor of five. That 
implies that the chemical processes responsible for the amides generation are  established early in molecular cloud evolution and are not altered during the star-forming processes, suggesting that to better understand amides formation, attention should be directed toward cold cloud conditions.
In such environments, amides formation might take place on the surface of 
interstellar icy-grains\cite{bisschop_testing_2007,lopez-sepulcre_shedding_2015}.
Moreover, the constant column density ratios between the species suggests that similar, or linked, chemical 
pathways might  be  responsible for the formation of the observed amides. 
Carbamoyl radical (\ce{NH2CO}) is considered a common precursor to interstellar amides\cite{hubbard_ultraviolet-gas_1975,agarwal_photochemical_1985,ligterink_formation_2018,chuang_formation_2022}. 
This particular species has received considerable attention due to its involvement in various straightforward synthetic pathways employed for the generation of amides.
Starting from carbamoyl, formamide  might be produced either by hydrogenation (\ref{eq:1}), 
disproportionation
 (\ref{eq:2}) or reaction with water  (\ref{eq:3}), however, the feasibility of Reaction \ref{eq:1} has long been debated\cite{noble_hydrogenation_2015,haupa_hydrogen_2019}:

\begin{equation}
 \ce{NH2CO} + \ce{H} \longrightarrow  \ce{NH2CHO}
\label{eq:1}
\end{equation}
\begin{equation}
 \ce{2NH2CO} \longrightarrow \ce{NH2CHO} + \ce{HNCO}
\label{eq:2}
\end{equation}
\begin{equation}
 \ce{NH2CO} + \ce{H2O} \longrightarrow \ce{NH2CHO} + \ce{OH} 
 \label{eq:3}
\end{equation}

Addition  of \ce{CH3} and \ce{NH2} to the carbamoyl radical leads to the formation of 
acetamide (\ref{eq:4}) and urea (\ref{eq:5}), respectively:

\begin{equation}
  \ce{NH2CO} + \ce{CH3} \longrightarrow  \ce{CH3C(O)NH2} 
\label{eq:4}
\end{equation}
\begin{equation}
\ce{NH2CO} + \ce{NH2} \longrightarrow \ce{NH2C(O)NH2} 
\label{eq:5}
\end{equation}

\noindent Reaction \ref{eq:4} has been proposed\cite{ligterink_prebiotic_2022} as a potential
reaction pathway linking the formation of formamide and acetamide, but the process has not been investigated so far. On the other hand, both experimental\cite{ligterink_formation_2018} and computational\cite{slate_computational_2020} studies have identified Reaction \ref{eq:5} as a particularly promising pathway.

The generation and availability of carbamoyl radical \ce{NH2CO} in the ISM is still object of study. 
Possible formation pathways\cite{belloche_rotational_2017,belloche_re-exploring_2019,garrod_formation_2022} are radical addition of \ce{NH2} to CO (\ref{eq:6}), 
H-abstraction from formamide (\ref{eq:7}), 
hydrogenation of HNCO (\ref{eq:8}), and hydration of CN (\ref{eq:9}):
\begin{equation}
\ce{NH2} + \ce{CO} \longrightarrow \ce{NH2CO} 
\label{eq:6}
\end{equation}
\begin{equation}
\ce{NH2CHO} \longrightarrow \ce{NH2CO}  + \ce{H}  
\label{eq:7}
\end{equation}
\begin{equation}
\ce{HNCO} + \ce{H} \longrightarrow \ce{NH2CO} 
\label{eq:8}
\end{equation}
\begin{equation}
\ce{CN} + \ce{H2O} \longrightarrow \ce{NH2CO} 
\label{eq:9}
\end{equation}

\noindent Reaction 
\ref{eq:6} has been suggested based on experimental results
\cite{hudson_new_2000,bredehoft_electron-induced_2017,chuang_formation_2022},
in which different ice-mixtures of CO:\ce{NH3} and 
\ce{H2O}:CO:\ce{NH3} 
are exposed to  vacuum ultraviolet photons in an ultra-high vacuum chamber at 10 K.
\citeauthor{chuang_formation_2022}\cite{chuang_formation_2022} found that the 
highest yields of formamide are obtained when including \ce{H2O}  in the ice mixture, suggesting an important role  of water in facilitating the formation reactions.
On the other hand, the route involving HCN/CN has also been experimentally probed by energetic processing of 
HCN:\ce{H2O} ice \cite{gerakines_ultraviolet_2004} in 
which formamide and HNCO were identified. This 
reaction pathway has also been studied theoretically using a a 33-water molecules cluster as ice model\cite{rimola_can_2018}.

\begin{figure}[h]
\centering
\includegraphics[width=\textwidth]{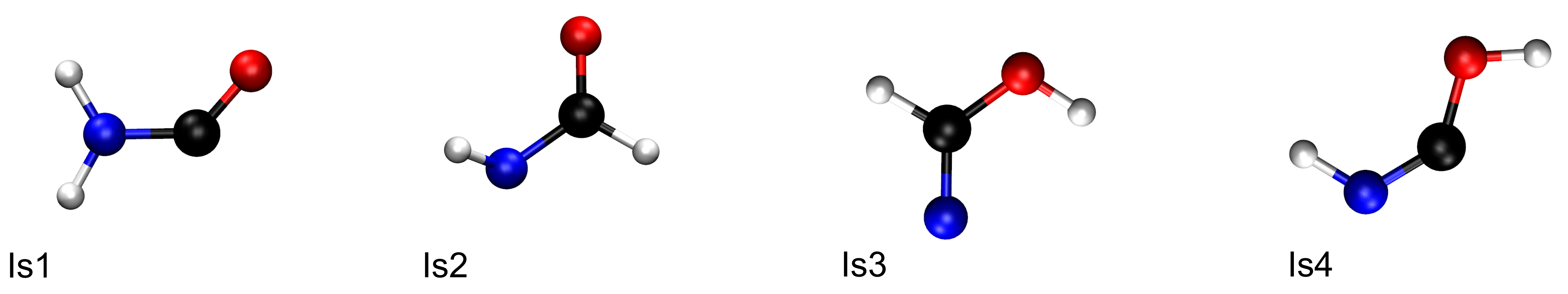}
\caption{Carbamoyl radical, Is1 and three of its isomers, Is(2-4). The color scheme for the atoms is red for O, black for C, blue for N and
white for H.}
\label{fig:isomers}
\end{figure}

Carbamoyl radical also presents several isomers, reported in Fig. \ref{fig:isomers}, 
although it is unclear how these species differ in term of reactivity and stability. 
Among them, HN(C=O)H, (Fig. \ref{fig:isomers}, Is2), whose addition to   \ce{CH3} radical leads to the formation of    
N-methylformamide (\ref{eq:10}): 

\begin{equation}
\ce{HN(C=O)H} + \ce{CH3}\longrightarrow \ce{CH3NHCHO}\\
\label{eq:10}
\end{equation}

\noindent Analogously to carbamoyl,  HN(C=O)H  can be formed
via radical recombination (NH + HCO), or H-abstraction from formamide\cite{belloche_rotational_2017,ligterink_formation_2018,garrod_formation_2022}.

When adsorbed on ice mantles, carbamoyl isomer are thought to convert between each other via tautomerization, so that potentially each of them 
could act as 
intermediate in the formation of several amides, as well. The tautomerization (\ref{eq:11}) of \ce{H(N=C)OH} isomer to carbamoyl (Is4 to Is1 respectively) has been investigated using an ice model
\cite{rimola_can_2018}, but the barrier for the reaction  was found to be too high to take place in the coldest regions on the ISM,
unless the 
process takes place following previous exothermic reactions, such that it might  absorb the energy
released by them:

\begin{equation}
 \ce{H(N=C)OH} \longrightarrow \ce{NH2CO}
 \label{eq:11}
\end{equation}

Interstellar ices are primarily composed of water in amorphous form (ASW)\cite{hama_surface_2013}; interstellar species interact with ASW in different ways,  due to the  
complex hydrogen bond (H-bond) network present on the ice surface. 
The binding energy (BE) of a species is a measure of the strength of its interaction with the surface, and can provide information about its residency time and consequent availability to reactive encounters with other species. According to experimental and theoretical studies, the collection of BE for a certain species, considering the most complete set of binding sites on the ice, adds up to constitute a Gaussian BE distribution\cite{he_interaction_2011,noble_thermal_2012,grassi_novel_2020,bovolenta_high_2020}.     
The ASW H-bond network may also affects surface processes, enabling or disfavoring certain chemical reactions in specific ways.  
Therefore, adopting a realistic ice model is essential to gain insight 
on the surface
formation routes of carbamoyl, and its isomers,
and consequently 
elucidating the origin of interstellar amides.\\

A pathway not yet explored for the generation of carbamoyl on the ASW surface, is the
addition reaction of hydroxyl radical (OH) to the common interstellar precursor HCN:
\begin{equation}
 \ce{HCN}+ \ce{OH} \longrightarrow \ce{HC(OH)N}\\
\label{eq:hcn_oh}
\end{equation}
\noindent Reaction \ref{eq:hcn_oh} yield the imidic acid isomer  of carbamoyl (\ce{HC(OH)N}, Fig. \ref{fig:isomers}, Is3). Gas-phase studies for this reaction have shown that the addition pathway of OH radicals 
to HCN has significantly lower barrier than the competitive H-abstraction\cite{galano_mechanism_2007,bunkan_theoretical_2013}:
\begin{equation}
 \ce{HCN}+ \ce{OH} \longrightarrow \ce{CN} + \ce{H2O}
\label{eq:hcn_oh_2}
\end{equation}

\noindent When OH is added to  \ce{CH3CN} species instead, the reaction  yields the corresponding acetamide precursor radical:

\begin{equation}
 \ce{CH3CN}+ \ce{OH} \longrightarrow \ce{CH3C(OH)N}\\
\label{eq:ch3cn_oh}
\end{equation}

\noindent Moreover, OH radicals can be easily produced in interstellar conditions, not only by the
photolysis and/or ion bombardments of \ce{H2O} but also the reaction of H and O atoms\cite{garrod_three-phase_2013,tsuge_behavior_2021}, so that they are thought to be one of the most abundant radicals on ice dust\cite{miyazaki_direct_2022}.\\

In this work, 
we present an estimation of the BE range and relative  energetic stability of the census of the 
isomers of carbamoyl radical, adsorbed on a minimal ice model, constituted of clusters of up to 5 water molecules each. 
We propose the first exploration of the HCN and \ce{CH3CN} + OH addition pathways (Reactions \ref{eq:hcn_oh}, \ref{eq:ch3cn_oh}) 
on a ASW surface model composed of 22-water molecules, 
as a synthetic route to \ce{NH2CO} and \ce{CH3C(OH)N} amides precursors. 
The study is carried out by means of Density Functional Theory (DFT), using the BE 
distribution of HCN and \ce{CH3CN} adsorbed on ASW, as a starting point to select possible catalytic 
sites for the reaction.
The prospective binding sites of HCN and \ce{CH3CN} are sampled with their reaction partner OH, in order to    
to find viable reaction pathways.
Lastly, we applied the same protocol to the analysis   
of the most exploited surface synthetic routes for the production of carbamoyl isomers, namely the radical addition of \ce{NH2} to CO and of CN to \ce{H2O} (Reactions \ref{eq:6} and \ref{eq:7}, respectively), as well. 
 \\

\section{Computational details}\label{sec2}
\subsection{ASW modeling}
\begin{figure}[h]
\centering
\includegraphics[width=0.9\textwidth]{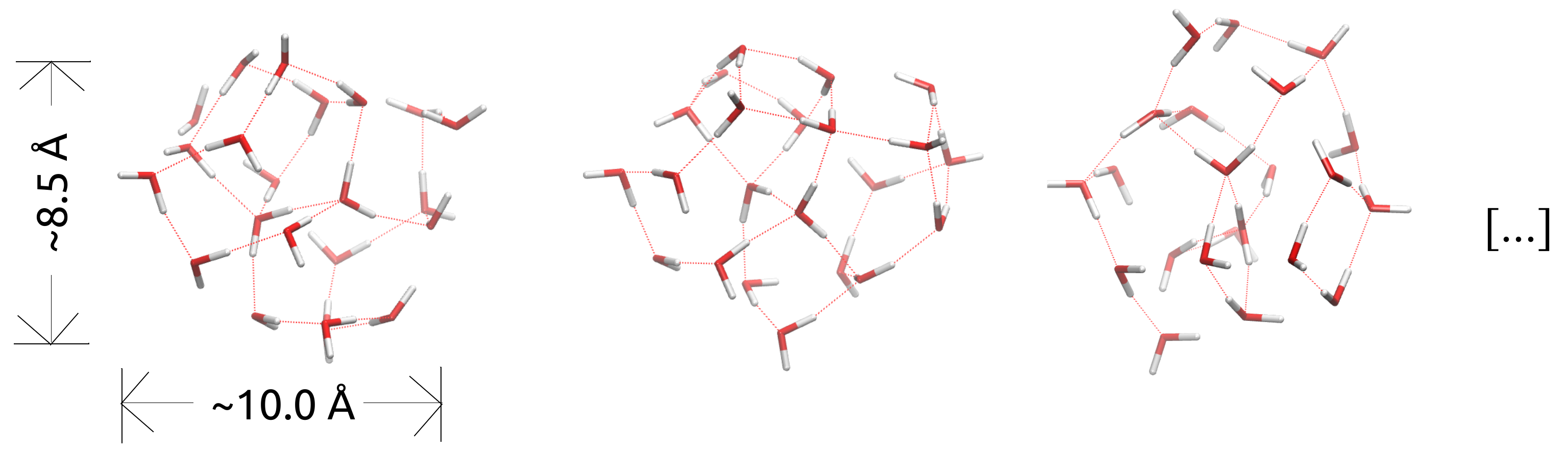}
\caption{This work utilizes a selection of 17 homogeneous amorphous clusters, each containing 22 water molecules, for evaluating binding energies and investigating reactivity. Following modeling using AIMD techniques, the structures undergo geometry optimization. The color scheme for the atoms is red for O  and
white for H.}
\label{fig:clusters}
\end{figure} 
We used the 22-water molecules set-of-clusters model, which is part 
of the binding energy evaluation platform\cite{bovolenta_binding_2022} (BEEP) to find suitable binding sites 
for the addition reactions. This model consists of 17 homogeneous amorphous clusters
that are extracted from an \textit{ab initio} MD simulation of a singular cluster 
containing 22 water molecules. Three examples of the clusters are shown in Fig. \ref{fig:clusters}. The details about the simulation procedure can 
be found elsewhere\cite{bovolenta_high_2020}. As  each  of the 17  clusters 
represent unique fragments of ASW surface, the set-of-cluster encompass 
a rich variety of binding sites, thus providing many possible sites 
that could be susceptible to reactive encounters.  
\subsection{Binding Energy and Binding Energy Distributions}
The binding energy of a species ($i$) adsorbed on a surface ($ice$) is defined as:    
\begin{equation}
BE(i) = E_{sup} - (E_{ice}+ E_i)
\label{eq:BE}
\end{equation}
where $E_{sup}$ stands for the energy of the supermolecule formed by the adsorbate bound to the surface, $E_{ice}$ refers to the surface energy, and $E_i$ is the energy of the adsorbate.
The BE is assumed to be a positive quantity, according to convention. 
Using a BE distribution of values reflects 
a more realistic desorption behavior for molecules adsorbed on ASW ice. 
The binding energy 
distributions of \ce{HCN} and \ce{CH3CN} are computed using 
using BEEP computational platform and protocol\cite{bovolenta_binding_2022}\footnote{https://github.com/QCMM/BEEP}. 
\hg{The performance of DFT largely depends on the exchange-correlation density functionals employed. Certain functionals are tailored for specific systems or optimized to accurately reproduce geometries or energy barriers. Additionally, these functionals differ significantly in computational cost. Common practice involves testing (benchmarking) a range of DFT functionals and basis sets on minimal surface models to identify the most suitable method for application to larger systems\cite{shingledecker_chapter_2024}.}
Therefore, the  level of theory selected in this study is based on a extensive DFT geometry 
and energy benchmark carried out in a previous work\cite{bovolenta_binding_2022}. 
The geometries  of the binding sites have been computed using HF-3c method coupled with MINIX basis set\cite{sure_corrected_2013}
while 
MPWB1K\cite{zhao_hybrid_2004} coupled with def2-TZVP\cite{weigend_balanced_2005}
is used for the binding energies.

Dispersion effects are treated using D3BJ correction\cite{grimme_effect_2011}.
\textsc{TeraChem}\cite{seritan_terachem_2021} and \textsc{Psi4}\cite{turney_psi4_2012} softwares are used for 
the optimization of the binding sites and \textsc{Psi4} for binding energy computations.
BEEP is powered by the QCArchive\cite{smith_molssi_2021} software stack.

\subsection{PES exploration with QCExplore}
For the exploration of the PES we employed our newly developed 
QCExplore platform that automatizes the search for possible reactive 
complexes on the ASW surface and streamlines the search for transition 
state structures. It consists of four pillars: prospective reactive binding
site sampling with target molecule, transition path search by means of relaxed 
scan or nudge elastic band calculations, transition state characterization, by 
computing the Hessian and analyzing the normal mode frequency and  
displacing and relaxing the TS geometry along the reactive normal modes with imaginary
frequency, to corroborate that the TS corresponds to the correct reactants and products. 
Finally, the workflow also allows to compute the reaction and transition state energies 
at various DFT levels of theory. QCExplore is powered by the QCArchive software 
stack, as well.  The source code of QCExplore can be found on GitHub\footnote{https://github.com/QCMM/QCExplore}. Within QCExplore, we use 
\textsc{Terachem} together with geomeTRIC\cite{wang_geometry_2016} for the geometry optimizations of reactants and
transition states using BHandHLYP\cite{becke_new_1993} and def2-SVP. The relaxed scan and NEB computations are also  done with 
\textsc{Terachem}/geomeTRIC. Finally the energies of the stationary points are 
computed with Psi4 using BMK\cite{boese_development_2004} coupled with def2-TZVP basis, including D3BJ correction.

\subsection{Sampling of prospective binding sites.}\label{sec:sampling}
The procedure to identify the reactant configuration for the reactions carried out on ASW is the following:
\begin{itemize}
\item Selection of prospective binding sites of one of the reactive partners adsorbed on ASW (HCN for R1, \ce{H2O} for R2,  \ce{NH2} for R3 and \ce{CH3CN} for R4)      
\item Extensive sampling with the other fragment: the sampling procedure creates a large number (10-15) of potential reaction sites, where the fragment is  
arranged on a semi-spherical grid around the adsorbed species. Each generated configuration 
is optimized and a suitable reactant complex is chosen to study the reaction. The radius of the hemisphere is 
3 \AA. 

\end{itemize}

\section{Results and Discussion}\label{sec:results}
This section is organised as follow: 
first, we present the analysis of the structural parameters and relative energies of carbamoyl isomers (Sec.\ref{sec:carb}) and their BE range (Sec. \ref{sec:carb_BE}).
Then, we present the reactivity results 
for the generation of \ce{NH2CO} radicals,  
in gas-phase and on  
ASW for the Reaction of OH + HCN (Sec.\ref{sec:r1_gf},\ref{sec:r1_asw}), CN + H2O (Sec. \ref{sec:r2}), NH2 + CO (Sec. \ref{sec:r3}). Finally, we describe the results of the reaction of OH + \ce{CH3CN} (Sec. \ref{sec:r4}).    
The BE distribution and binding mode analysis 
of HCN and \ce{CH3CN} adsorbed on ASW are displayed at the beginning of the corresponding sections (Sec. \ref{sec:r1_asw}, \ref{sec:r4}).

\subsection{Structural parameters and relative energies of carbamoyl isomers} \label{sec:carb}
We selected the four most common radical isomers with formula \ce{NH2CO}, 
Table \ref{tab:bonddist_CCSDT} reports their structural parameters. The species share the same backbone, 
and differ based on the  position of the H-atoms. Is1 (C-radical carbamoyl, or simply carbamoyl) and Is2 (N-radical carbamoyl) present a 
carbonyl group, which is reflected in the shorter distance of the C--O bond (B(N--C) is 1.20, 1.23 \AA, respectively). 
Is2 is the most strained isomer, with a N-C-O angle value of 123.7 $^o$.
Is3 is 
a imidic acid, characterized by the presence of both a C=N bond and a OH group. Lastly,
Is4 represents the straight-chain isomer, as indicated by the largest N-C-O angle value (132.5 $^o$). In its lowest energy conformation, the N--H and O--H 
groups in Is4 are pointing in opposite directions.\\
\begin{table}[h]
\centering   
\caption{Structural parameters of carbamoyl (Is1) and its isomers (Is2-4). Geometries are obtained at UCCSD(T)-F12/cc-pVDZ-F12 level of theory. Bond distances (B) are in angstrom (\AA), bond angles (A) are in degrees ($^o$).} 
\begin{tabular}{c llll}
\hline
          Isomer  & B(N--C) & B(C--O) & A(N-C-O)\\
          \hline
         Is1 & 1.34 & 1.20  & 129.6 \\
         Is2 & 1.38 & 1.23  & 123.7 \\
         Is3 & 1.26 & 1.38  & 125.5 \\
         Is4 & 1.23  & 1.35  & 132.5 \\
\hline
\end{tabular}
\label{tab:bonddist_CCSDT}
\end{table}

\begin{figure}[H]
\centering
\includegraphics[width=0.6\textwidth]{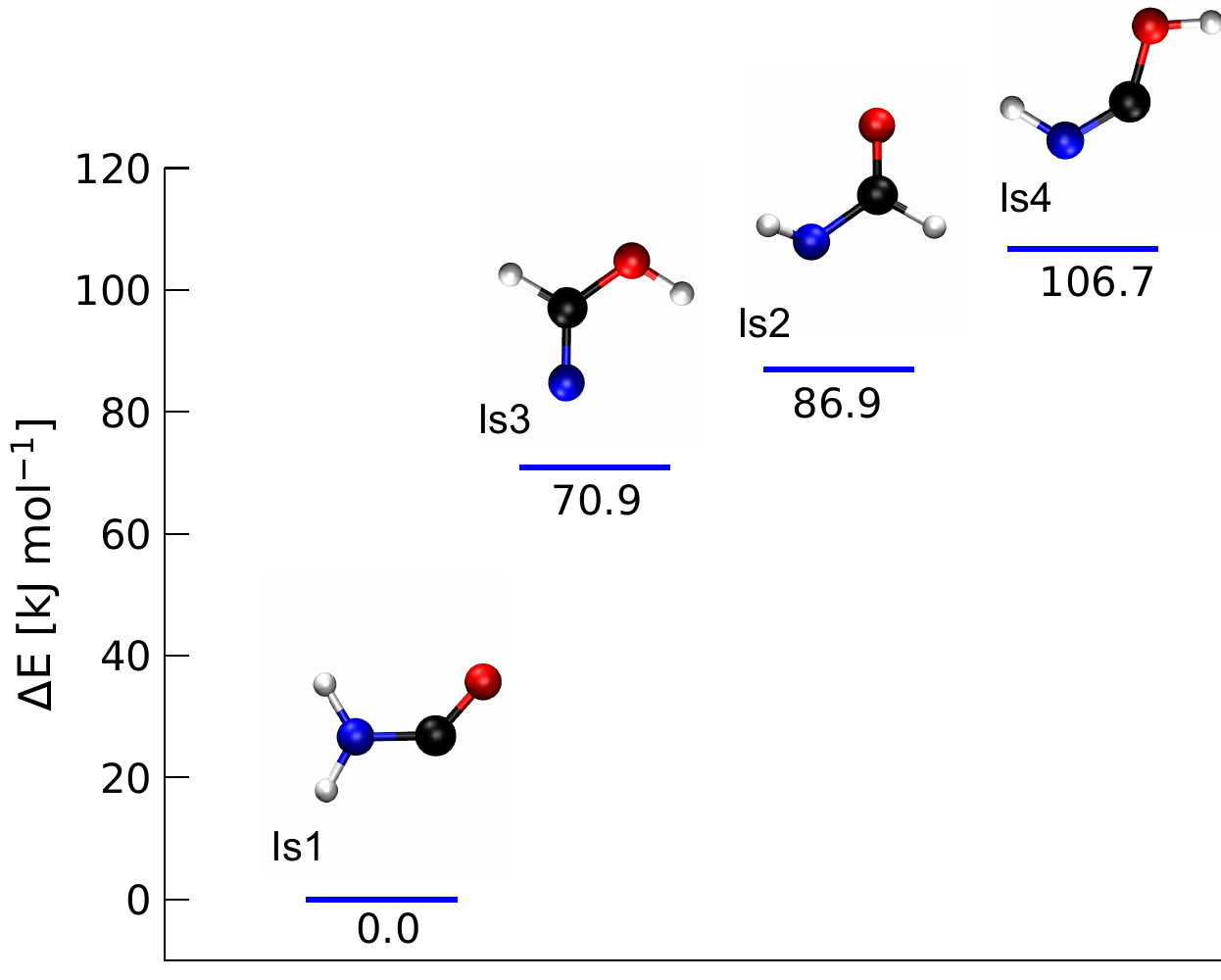}
\caption{Relative energy of the radical isomers studied in  this work, with respect to carbamoyl (Is1). Energies and geometries are computed at UCCSD(T)-F12/cc-pVDZ-F12\cite{warden_efficient_2020} level of theory.}
\label{fig:en_isomers}
\end{figure} 
Fig. \ref{fig:en_isomers} display the relative energies of the different isomers with respect to carbamoyl (Is1).  The following isomers in order of stability appears to be the imidic acid (Is3), and the N-radical form of the amide radical (Is2) with a energy difference of 70.9 and 86.9 kJ/mol, respectively.
The least stable species is the linear iminic acid isomer (Is4), 106.7 kJ/mol higher in energy than carbamoyl.  

\subsection{Binding energy range of carbamoyl isomers}\label{sec:carb_BE}
\begin{figure}[H]
\centering
\includegraphics[width=\textwidth]{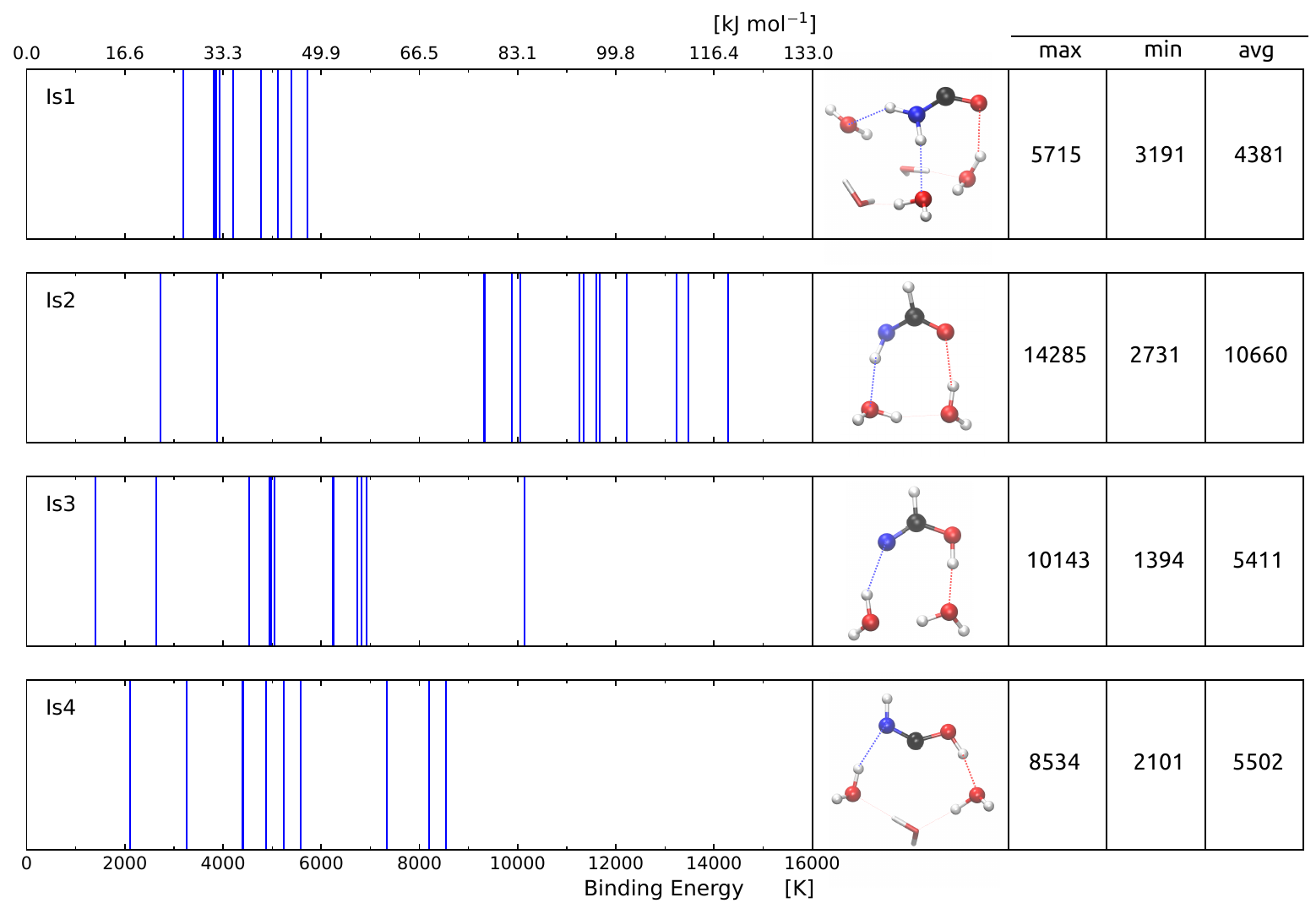}
\caption{
\hg{Range of BE calculated in this work for carbamoyl isomers (from top to bottom: Is1--4) adsorbed on a set of small water clusters, \ce{(H2O)_{1--5}}. The energy has been computed at $\omega$-PBE-D3BJ/def2-TZVP  
level of theory, using PWB6K-D3BJ/def2-TZVP geometries. The  first column reports the set of BE values obtained for a specific isomer: system optimization and BE calculation are done individually for each binding site, hence the blue lines refer to the BE of a specific \ce{Is(1-4)}--\ce{(H2O)_{1--5}} structure. The second column reports the structure that provides the highest BE of the set, for all systems. The isomers and the water molecules to whom they are directly interacting with, are represented as balls and sticks, the rest as sticks. The H-bond interactions established by the isomers are highlighted. The remaining columns report the maximum (max), minimum (min) and average (avg) BE values obtained for each set, in K.}}
\label{fig:be_carb}
\end{figure} 
In order to estimate the BE of carbamoyl and the other isomers absorbed on ice, we considered 
a set of minimal water clusters (up to 5 water molecules, \hg{ \ce{(H2O)_{1--5}}). Sampling of the 
radicals around the clusters, followed by filtering according to geometrical criteria, provided a 
set of unique binding sites(\ce{Is(1-4)-(H2O)_{1--5}}), each of them entailing a specific 
BE value. Fig. \ref{fig:be_carb} reports the range of BE calculated for each isomer,  along with a 
image of the structure that provides the highest BE of the set.} \hg{The largest BEs appear to belong to isomer 
Is2, with a average BE of around 10600 K. In this isomer there is a possible resonance 
between the  carbonyl and iminic form, in which the unpaired electron is localized on the oxygen. 
Therefore, the combination of the  unpaired electron on the oxygen-end  acting as H-bond acceptor 
and the imine N--H moiety of the molecule acting as H-bond donor, provides strong adsorption of the species on ice}.
The isomers that present an alcoholic group, Is(3,4), also display similar BEs, \hg{centered around 5500 K}. 
Taking into account the structures with the highest BE \hg{for these isomers}, it can be noticed 
that the O--H group acts primarily as H-bond donor, while N--H group (in Is4) and the amidic group 
(in Is3) as H-bond acceptor, meaning that Is(3,4) also present similar adsorption motives.     
Finally, carbamoyl (Is1) is the radical with the lowest average BE 
(\hg{4381} K). However, analysis of the structures' geometries revealed that, in the majority of 
them, the binding \hg{does not take place through the C-atom, where the radical is located, 
which might provide a explanation for the range of lower BE values.}

\subsection{Formation pathways of carbamoyl isomers}

\begin{scheme}[h]
\centering
\includegraphics[width=0.6\textwidth]{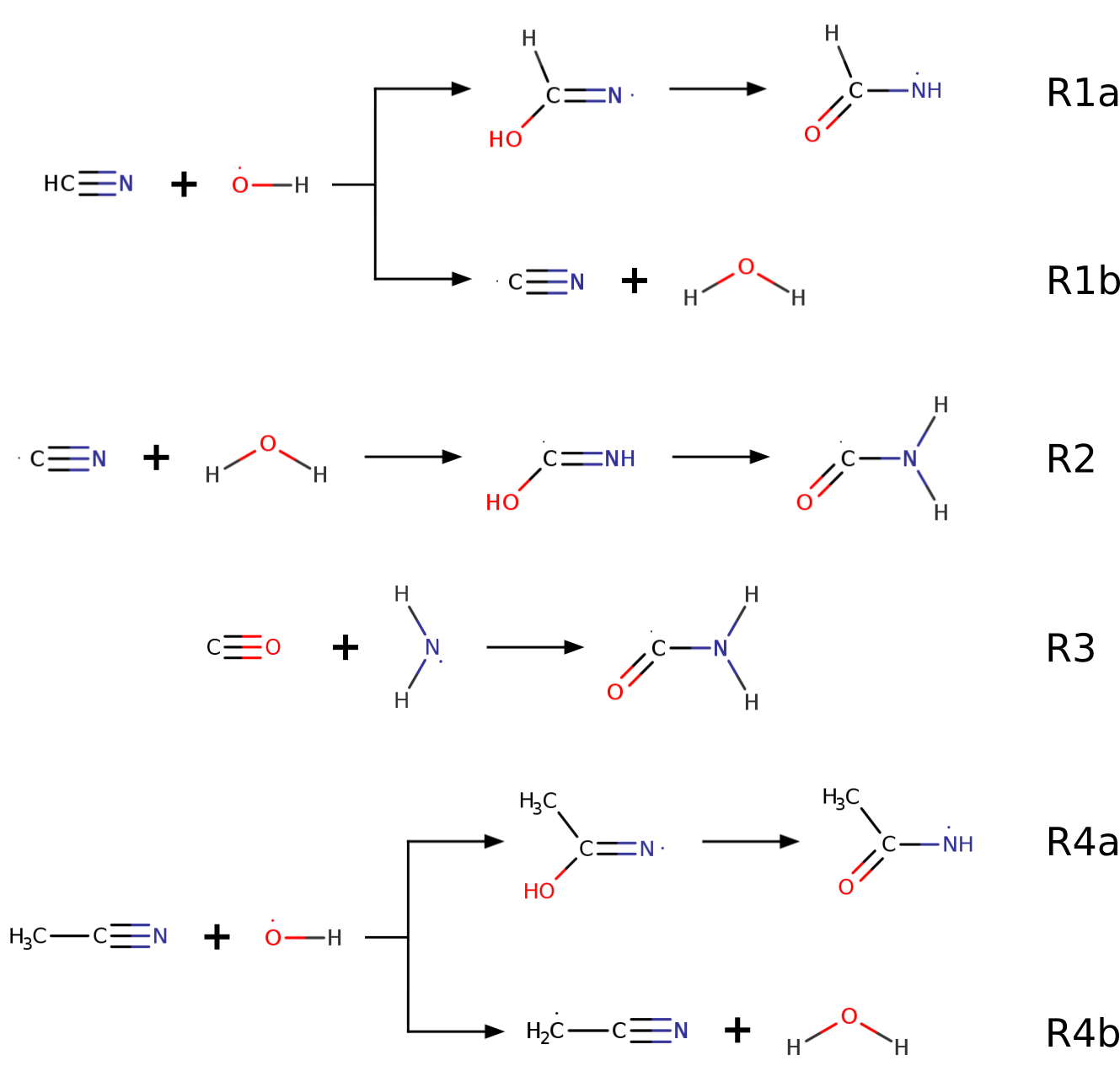}
\caption{Reactions pathways studied in this work.}
\label{sch:2d_scheme}
\end{scheme}

\begin{table}[]
\centering
\small
\begin{tabular}{ccccccccccc}
\hline
\multicolumn{3}{l}{\multirow{2}{*}{}} &
  \multicolumn{2}{c}{Addition} &
  \multicolumn{2}{c}{N-H Rotation} &
  \multicolumn{2}{c}{Tautomerization} &
  \multicolumn{2}{c}{H-Extraction} \\
\multicolumn{3}{l}{} &
  $\Delta E^{\ddagger}$ &
  $\Delta{E}^o$ &
  $\Delta E^{\ddagger}$ &
  $\Delta{E}^o$ &
  $\Delta E^{\ddagger}$ &
  $\Delta{E}^o$ &
  $\Delta E^{\ddagger}$ &
  $\Delta{E}^o$ \\ \hline
\multirow{2}{*}{R1} & \multirow{2}{*}{OH + HCN}   & Gas-phase & 15.6 & -130.1 & ---  & ---   & 165.1 & 8.3   & 82.7 & 42.8  \\
                    &                             & ASW       & 26.7 & -113.8 & ---  & ---   & 68.6  & 17.3  & 94.2 & 3.1   \\
\multirow{2}{*}{R2}     &   \multirow{2}{*}{CN + \ce{H2O}}         &   \hg{Gas-phase}\textsuperscript{\emph{a}} &  172.2 & --- & 130.1  & --- &  190.8 &  & ---  & ---   \\     
&     & ASW       & 70.6 & -81.9  & 38.5 & -36.0 & 6.2   & -84.5 & ---  & ---   \\
  \multirow{2}{*}{R3}            & \multirow{2}{*}{\ce{NH2} + CO}            &   \hg{Gas-phase}     & 10.2 & -120.0 & ---  & ---   & ---   & ---   & ---  & ---   \\        
 & & ASW       & 12.6 & -110.8 & ---  & ---   & ---   & ---   & ---  & ---   \\
\multirow{2}{*}{R4} & \multirow{2}{*}{OH + \ce{CH3CN}} & Gas-phase & 19.1 & -115.9 & ---  & ---   & 154.6 & 6.6   & 32.9 & -80.2 \\
                    &                             & ASW       & 40.2 & -105.1 & ---  & ---   & 59.2  & 6.9   & 32.0 & -76.9 \\ \hline
\end{tabular}
\textsuperscript{\emph{a}} \footnotesize {\hg{Values from the literature\cite{rimola_can_2018}, computed at BHYLP/6-311++G(d,p) level of theory and basis. }}
\caption{Energy barriers ($\Delta E^{\ddagger}$) and reaction energies ($\Delta{E}^o$) computed for all of the reactions studied, at BMK-D3BJ/def2-TZVP level of theory, using BHandHLYP-D4/def2-SVP geometries. Values in kJ $\mathrm{mol^{-1}}$.}
\label{tab:bigbigtable}
\end{table}

\begin{figure}[H]
\centering
\includegraphics[width=\textwidth]{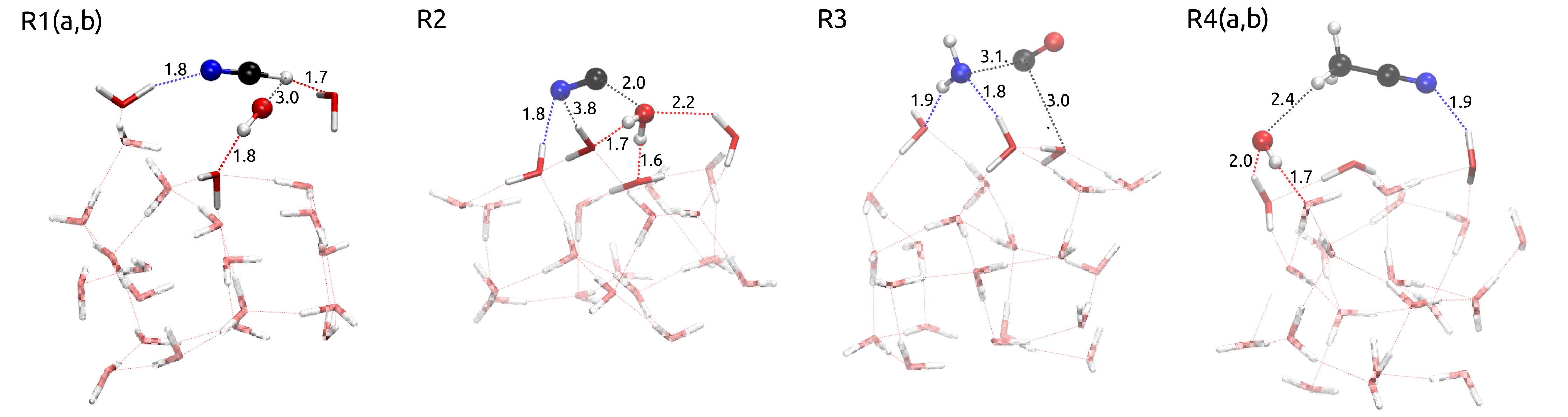}
\caption{Reactive sites selected for the processes studied in this work. Geometries obtained using BHandHLYP-D4/def2-SVP. \hg{The fragments that participate in the reaction are represented as balls and sticks, the rest as sticks. The water molecules that establish H-bond interactions with the reactants are highlighted. Relevant bond distances are reported in Å.} The color scheme for the atoms is red for O, black for C, blue for N and
white for H.}
\label{fig:reac}
\end{figure}

Scheme \ref{sch:2d_scheme} shows the reaction pathways studied in this work.
Each of them have been simulated using the ASW ice model, while R1, \hg{R3} and R4 are studied in gas-phase, as well.
For Reactions R1 and R4 we investigated two different outcomes, based on the addition reaction (a) and water elimination (b).  
Fig. \ref{fig:reac} shows the reactive site selected for R(1--4) on ASW. Table \ref{tab:bigbigtable} reports the energy barriers and reaction energies for all the pathways. 

\subsubsection{R1: Gas-phase reaction pathways for OH + HCN}\label{sec:r1_gf}

\begin{figure}[H]
\centering
\includegraphics[width=\textwidth]{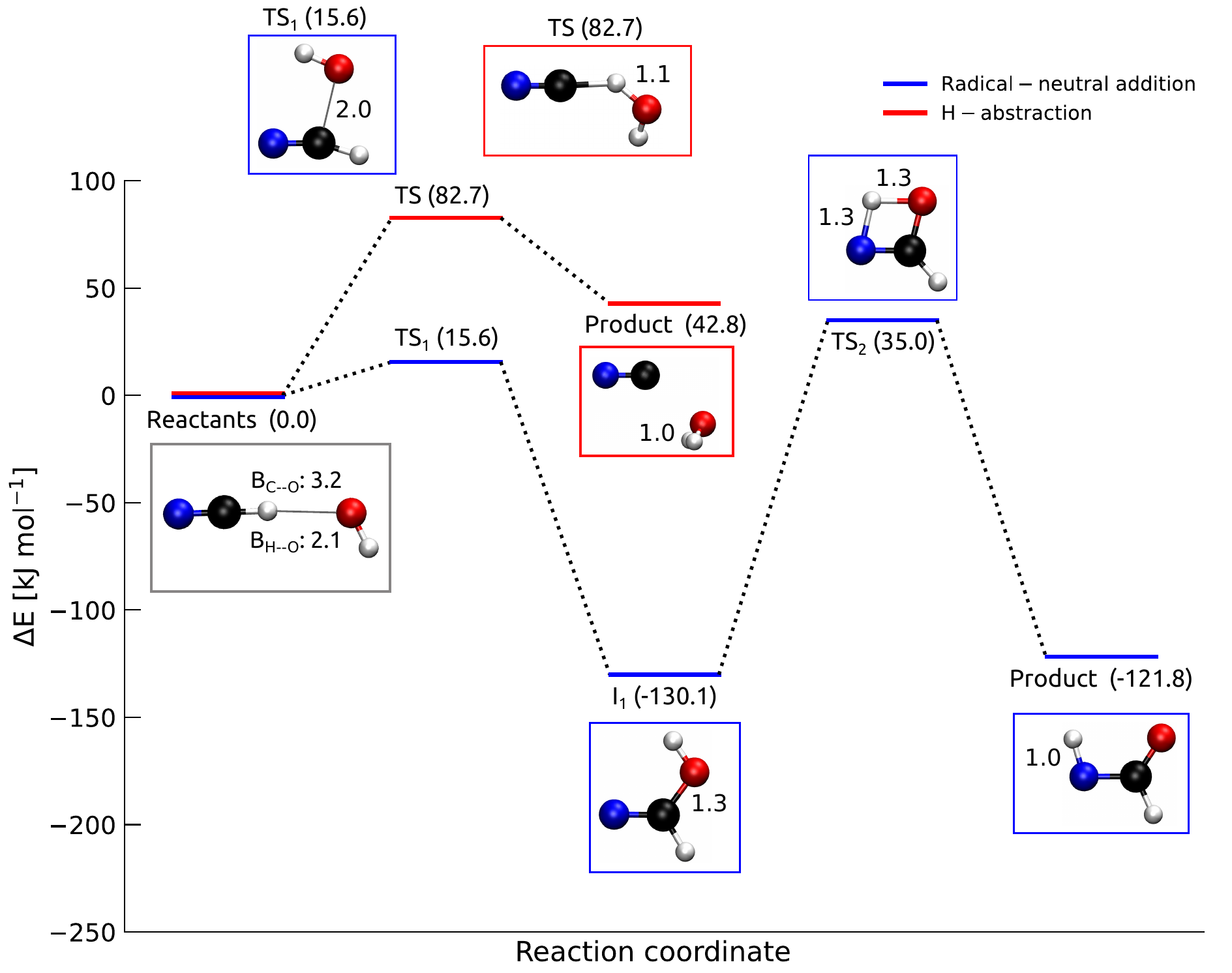}
\caption{Energy and stationary states for the gas-phase  reaction \hg{R1}:  OH + HCN, using BHandHLYP-D4/def2-SVP geometries and BMK-D3BJ/def2-TZVP
energies. The process has two possible outcomes: the addition, \hg{R1a} in blue, and the water elimination, \hg{R1b} in red.    Distances in Å. The color scheme for the atoms is red for O, black for C, blue for N and
white for H.}
\label{fig:hcn_oh_gf}
\end{figure} 

\noindent \textit{Radical-neutral addition: formation of Is3 and tautomerization from Is3 to Is2:}
Fig. \ref{fig:hcn_oh_gf} reports the energy diagram for the reaction.
The addition of OH to HCN (Fig. \ref{fig:hcn_oh_gf}, blue pathway) 
is composed of 
two 
steps: the addition step, which entails an energy barrier of  
15.6 kJ $\mathrm{mol^{-1}}$ and yields the intermediate imidic radical isomer Is3 (I1, Fig. \ref{fig:hcn_oh_gf}),  followed by the  tautomerization step
(energy barrier of 165.1 kJ $\mathrm{mol^{-1}}$) from isomer Is3 to Is2. It is important to note that for the tautomerization step in the 
gas-phase, a very strained configuration is needed for the transition state (\ce{TS_2}, Fig. \ref{fig:hcn_oh_gf}), resulting in a high TS energy. Furthermore,
the intermediate (Is3, methanimidic acid) corresponds to the thermodynamic product, while the amide form the radical (Is2) has slightly higher energy, consistent with the relative energies computed in 
Sec. \ref{sec:carb}.\\

\noindent \textit{H-abstraction: water elimination}
For the hydrogen abstraction (Fig. \ref{fig:hcn_oh_gf}, red pathway), the reaction presents a significant
barrier of 82.7 kJ $\mathrm{mol^{-1}}$, 67.1 kJ $\mathrm{mol^{-1}}$ larger than the barrier of the addition.
In light of this result, we conclude that 
the addition product is the most plausible pathway for this reaction channel in gas-phase.

\subsubsection{R1: ASW reaction pathways for OH + HCN}\label{sec:r1_asw}
\noindent \textit{Binding energy distributions and binding modes analysis}:

\begin{figure}[H]
\centering
\includegraphics[width=0.6\textwidth]{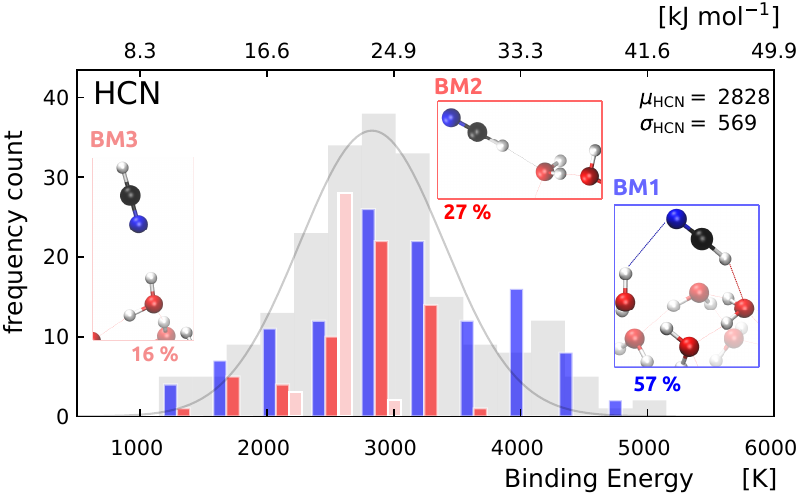}
\caption{Histogram of the BE distribution of HCN computed on a set-of-clusters model 
of 22 water molecules, calculated at MPWB1K-D3BJ/def2-TZVP//HF-3c/MINIX level of theory, without including
ZPVE correction. BE values are given in K (lower scale) and kJ $\mathrm{mol^{-1}}$
(upper scale). The figure displays a composite plot: the grey histogram refers to the total data, while the structures corresponding to different binding modes (BM1-3) are colored in blue, red and pink. 
Mean BE ($\mu$) and standard deviations of a Gaussian fit ($\sigma$) for the total distribution (grey data) are reported
in K. The insets show an
example of the different binding modes encountered.}
\label{fig:be_hcn}
\end{figure} 
Fig. \ref{fig:be_hcn} shows the BE distribution for HCN (upper panel) adsorbed on the set of ASW clusters,  in units of Kelvin 
and kJ $\mathrm{mol^{-1}}$. The distribution contains 212 unique binding sites and
displays a single marked distribution which is slightly asymmetric 
towards higher energies. The application of a Gaussian fit to the distributions allowed to estimate the mean ($\mu$) and standard deviation ($\sigma$). 
For HCN, $\mu$ is 2828 K with $\sigma$ of 569 K. We have been able to identify three binding modes, based on the analysis of characteristic distances in the adsorption sites. The binding mode that exhibits the highest BE values and also the most common (BM1, 57\%, in blue in Fig.
\ref{fig:be_hcn}) is given by the creation of two H-bonds via both the extremities of the molecule. The H-atom act as H-bond donor while the interaction on the side of N-atom is established by one of the surface molecules. 
The other binding modes present solely one of the H-bond interactions: HCN acting as H-bond donor (BM2, 27\%, red) and as H-bond acceptor (BM3, 16\%, pink), and show lower BEs, suggesting that both the interactions are paramount to produce strong binding sites. 
The latter modes consequently display a minor level of insertion into the ice network.\\      
In order for the addition of OH to HCN (R1a) to take place, a C--O bond needs to be established 
with the central C-atom of HCN, thereby, the backbone of the molecule need to be available for the reaction. Some of BM1 structures, being well inserted in the surface H-bond network, might present steric hindrance, preventing the reaction. On the other hand, in binding sites where the molecules are pointing upward away from the surface (BM2-3), the C-atom is located further away with respect to the surface, making it difficult to reach for a roaming radical.   
Regarding the H-abstraction from HCN (R1b), apart from BM3 binding sites where the H-extremity is available, the reaction requires the breaking of a H-bond interaction between the molecule and the surface, as well as the breaking of C--H bond, hinting that the process might be favored in sites where the HCN is loosely bound. 
Such considerations underline  
the importance of selecting the proper site for the reaction, as the ASW offers a large variety of binding situations, even for a small species like HCN.\\ 

\begin{figure}[H]
\centering
\includegraphics[width=\textwidth]{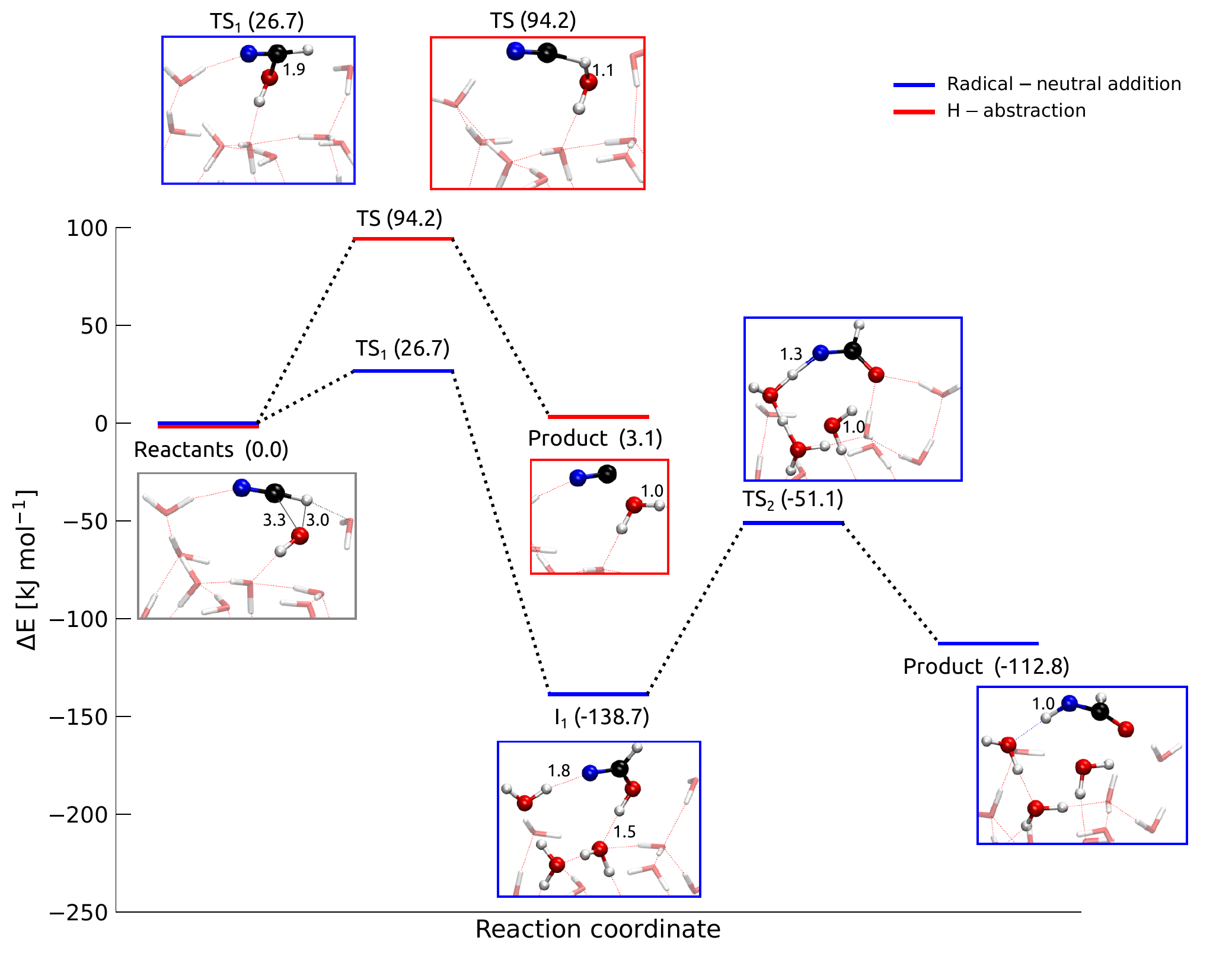}
\caption{Energy and stationary states for the ASW reaction \hg{R1}:  OH + HCN, using BHandHLYP-D4/def2-SVP geometries and BMK-D3BJ/def2-TZVP energies.
The process has two possible outcomes: the addition, \hg{R1a} in blue, and the water elimination, \hg{R1b} in red. Distances in Å. The color scheme for the atoms is red for O, black for C, blue for N and
white for H.}
\label{fig:hcn_oh_asw}
\end{figure} 

\noindent \textit{Radical-neutral addition: formation of Is3 and tautomerization from Is3 to Is2:}
\noindent 
Fig. \ref{fig:hcn_oh_asw} shows the energies and stationary states for the
OH + HCN for both the addition and hydrogen extraction reactions over the ASW surface. 
The reactant complex has been obtained following the procedure in Sec. \ref{sec:sampling}; as prospective binding site for HCN we selected a high BE structure, belonging to the BM1 binding mode, from the binding mode analysis  previously carried out. The reactive site is reported in Fig. \ref{fig:reac}: the HCN fragment is establishing H-bonds with two surface water molecules, via both its extremities.\\      
For the addition reaction, we also studied the possibility of a tautomerization 
reaction to transform the imidic-acid tautomer (Is3) to the amide radical (Is2). 
The OH + HCN addition reaction presents a TS energy of 26.7 kJ $\mathrm{mol^{-1}}$. 
The resulting imidic-acid intermediate is exothermic with an energy of -138.7 kJ $\mathrm{mol^{-1}}$. 
The tautomerization reaction from the imidic to the amidic radical (Is3 to Is2) presents a high TS barrier of 68 kJ $\mathrm{mol^{-1}}$.  
This surface tautomerization reaction involves a 
proton relay carried out by three surface water molecules, a type of 
mechanism that has been observed in other isomerization  and addition 
reactions on ASW surface models\cite{rimola_can_2018,woon_ab_2002, baiano_gliding_2022}. 
Furthermore, despite the high barrier, the TS energy remains below that of the
OH + HCN reactant complex, indicating that the energy released from the first 
highly exothermic step could potentially  be utilized to overcome this barrier. 
Finally, it is worth noting that the overall lowest energy state on this PES is 
the Is3, the imidic-acid radical isomer (Fig. \ref{fig:hcn_oh_asw},  \hg{\ce{I_1}}),  and the energy difference between the imidic-acid 
and the amide radical (Is2) is further increased compared to that in the gas-phase.

\noindent \textit{H-abstraction:}
The alternative H-abstraction pathway for OH + HCN
has a TS barrier of the 94.2 kJ $\mathrm{mol^{-1}}$, which is significantly higher than the 
addition barrier. 
On the ASW surface, the products of 
the H-abstraction reaction are \hg{slighly} endothermic.

\subsubsection{R2: ASW reaction pathway for CN + \ce{H2O}}\label{sec:r2}
\begin{figure}[H]
\centering
\includegraphics[width=\textwidth]{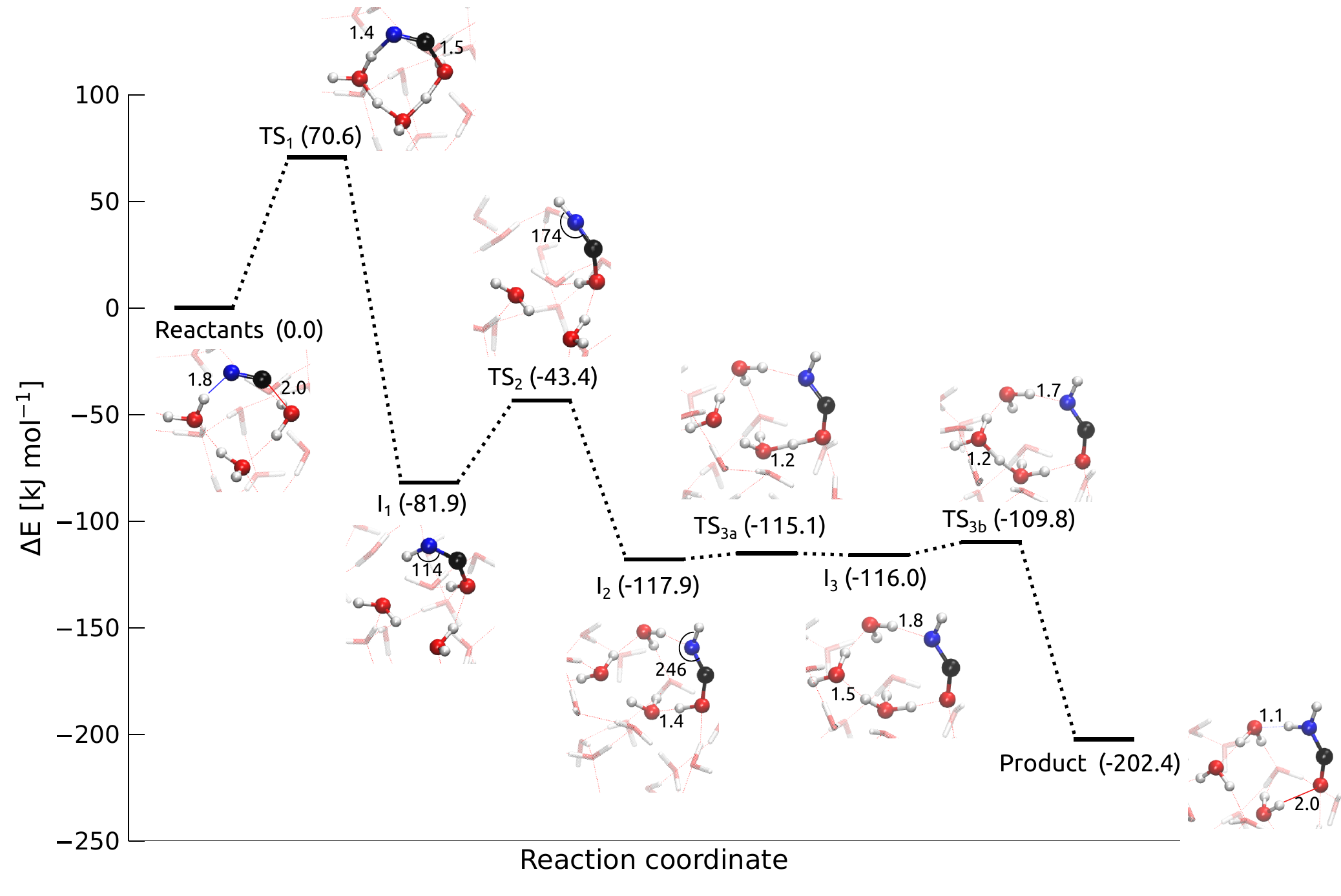}
\caption{Energy and stationary states for the ASW reaction \hg{R2}: CN + \ce{H2O}, using BHandHLYP-D4/def2-SVP geometries and BMK-D3BJ/def2-TZVP
energies. Distances in Å, angles in degree. The color scheme for the atoms is red for O, black for C, blue for N and
white for H.}
\label{fig:cn_w22}
\end{figure}

\noindent \textit{Radical-neutral addition: formation of Is4 and tautomerization to Is1:}
We studied the addition pathway of CN + \ce{H2O}. The reactive site structure is reported in Fig. \ref{fig:reac}, while the PES and the results in 
terms of stationary point geometries and energies are shown in Fig. \ref{fig:cn_w22}. 
The reactive site has been selected according to the binding mode analysis for \ce{H2O} molecule carried out in a previous work of our group \cite{bovolenta_binding_2022}. The binding site of \ce{H2O} is of high BE, as the molecule is strongly inserted into the H-bond network of the surface, as it can be noted by the high number of interactions that it is establishing with the neighbouring molecules (highlighted in Fig. \ref{fig:reac}).\\    
Differently to the reactions of OH with HCN, in this reaction 
the addition to the surface water molecule triggers a proton-relay  in 
the first phase of the reaction. This involves a hydrogenation of 
the N-atom end of the CN radical to yield the isomer Is4 (\hg{Fig. \ref{fig:cn_w22}, \ce{I_1}})  
and has therefore a significantly higher reaction barrier than the OH addition 
to HCN.  Furthermore, for the subsequent tautomerization reaction to take place, converting the Is4 isomer to the carbamoyl Is1,  
a second proton relay is necessary. The 
conformation of the first intermediate \ce{I_1}, however, is not oriented properly and in order for the tautomerization to proceed, the N--H group needs to rotate to establish an 
acceptor H-bond, instead  of a donor one.  The conformational change \hg{(\ce{TS_{2}})} presents a barrier of 38.5 kJ  $\mathrm{mol^{-1}}$, 
while the proton relay takes place in two steps \hg{(\ce{TS_{3a,b}})} and displays a very low total barrier of 6.2 kJ  $\mathrm{mol^{-1}}$, therefore,  the  
largest barrier in the tautomerization step corresponds to  
the conformational change in order to achieve the correct arrangement of the reactants.

The differences between our results and the work of \citeauthor{rimola_can_2018}\cite{rimola_can_2018}
, presenting a lower net TS barrier of 16.1 kJ  $\mathrm{mol^{-1}}$ 
for the addition and -74.8 kJ  $\mathrm{mol^{-1}}$ for the tautomerization, suggest 
that there might be a significant 
dependence of the TS energy on the binding site for this reaction. Furthermore, here we presents a different mechanism for the tautomerization, as that part of reaction CN + \ce{H2O} is step-wise, while in the work of \citeauthor{rimola_can_2018} it presents a concerted mechanism.
Moreover, in said study there was no need for a conformation change,  due to the particular choice in the 
binding site of CN on the ice model.  
The model used was obtained by fusing two smaller water clusters in order to form a cavity, therefore providing additional H-bond interactions to CN radical. Hence,
the peculiar binding of CN within such cavity allowed for two proton-relays to be carried 
out on the same binding site, since the N-atom was receiving two H-bonds in the reactive site. 
\subsubsection{R3: ASW reaction pathway for \ce{NH2} + CO}\label{sec:r3}

\begin{figure}[H]
\centering
\includegraphics[width=0.9\textwidth]{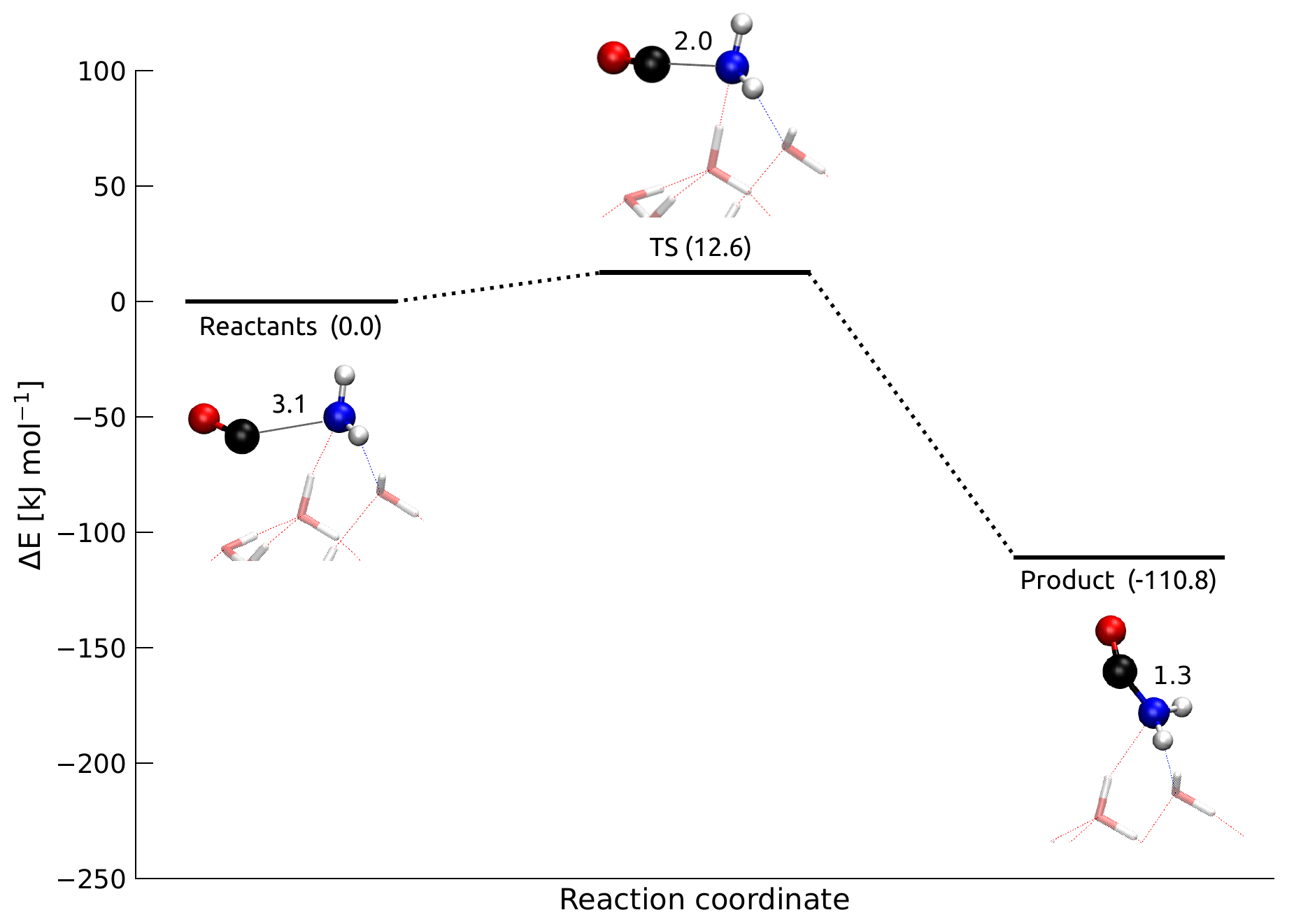}
\caption{Energy and stationary states for the ASW reaction \hg{R3}: \ce{NH2} + CO, using BHandHLYP-D4/def2-SVP geometries and BMK-D3BJ/def2-TZVP
energies. Distances in Å. The color scheme for the atoms is red for O, black for C, blue for N and
white for H.}
\label{fig:nh2_w22}
\end{figure} 

We studied the \ce{NH2} + CO addition on the ASW model surface, to form 
the carbamoyl radical (Is1). This reaction  is commonly included in astrochemical 
gas-grain models as a possible surface reaction to the formation of formamide\cite{garrod_three-phase_2013}. 
BE distribution  and binding mode analysis
for \ce{NH2} have been calculated in a previous work\cite{bovolenta_binding_2022}. A single binding mode was found, therefore we selected a high-BE site and sampled it with the CO molecule to find 
a potential reactant complex.  
Fig. \ref{fig:reac} reports the reactive site  selected to carry out the reaction on ASW. 
It can be seen that the \ce{NH2} radical forms two H-bonds with the water surface,
one being donated and one received, which determine the strong binding configuration for this species; 
while CO is not engaging in significant interaction
with the surface nor with \ce{NH2}.  

Figure \ref{fig:nh2_w22} shows the PES of this reaction.
The reaction is concerted and  leads to the formation of carbamoyl Is1. It presents a barrier of 12.6 kJ $\mathrm{mol^{-1}}$ \hg{(Fig. \ref{fig:nh2_w22}, TS)} and a marked exothermic character. This result corresponds to  the lowest barrier we encountered in this work for the formation of a carbamoyl isomer on ASW.  However, due to the  binding mode of the carbamoyl radical product,  further tautomerization to the imidic-acid form (Is3) could only be possible after a conformational change that establishes a receiving 
H-bond to the water surface through the carbonyl end.

\subsubsection{R4: Gas-phase reaction pathways for OH + \ce{CH3CN}}\label{sec:r4_gf}

Finally we probed a reaction pathway for acetamide by assessing the formation of 
the acetamide N-radical \ce{CH3(C=O)NH} precursor.

\begin{figure}[H]
\centering
\includegraphics[width=\textwidth]{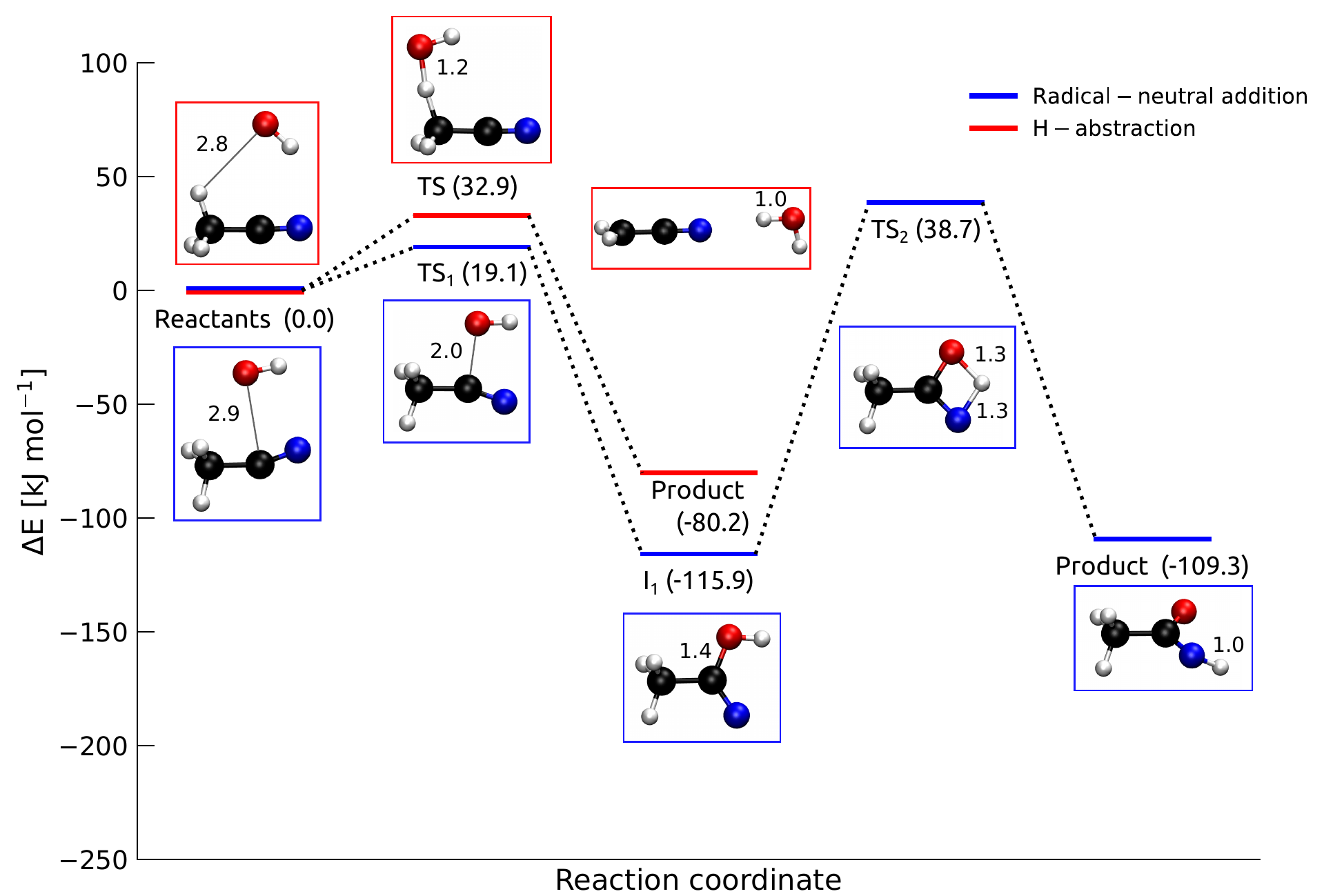}
\caption{Energy and stationary states for the gas-phase  reaction \hg{R4}:  OH + \ce{CH3CN}, using BHandHLYP-D4/def2-SVP geometries and BMK-D3BJ/def2-TZVP
energies. The process has two possible outcomes: the addition, \hg{R4a} in blue, and the water elimination, \hg{R4b} in red.    Distances in Å. The color scheme for the atoms is red for O, black for C, blue for N and
white for H.}
\label{fig:ch3cn_oh_gf}
\end{figure}

\noindent \textit{Radical-neutral addition: formation of the imidic isomer and tautomerization to the amidic form:}
The gas-phase addition of OH to \ce{CH3CN} (Fig.\ref{fig:ch3cn_oh_gf}, blue pathway) follows a similar mechanism to the addition of OH to HCN (R1a, Sec. \ref{sec:r1_gf}). 
The mechanism involved OH addition, giving the imidic radical, \ce{CH3C(OH)N},  with a barrier of 19.1 kJ $\mathrm{mol^{-1}}$, followed by the  
tautomerization to the amidic form: \ce{CH3(C=O)NH}, barrier 154.6 kJ $\mathrm{mol^{-1}}$.  
The barrier for the first step is slightly larger than that of reaction R1a, while the second 
step presents a smaller barrier by 10.5 kJ $\mathrm{mol^{-1}}$. The latter difference  might be attributed the fact that imidic acid intermediate is more constrained
than the equivalente state in the HCN + OH reaction pathway, 
and therefore is closer to \hg{\ce{TS_2}}, thus requiring less energy to reach it.

\noindent \textit{H-abstraction: water elimination}
The competitive H-abstraction (Fig.\ref{fig:ch3cn_oh_gf}, red pathway)
has a  barrier of 32.9 kJ $\mathrm{mol^{-1}}$, which is higher than the addition barrier of 
19.1 kJ $\mathrm{mol^{-1}}$. Hence, analogously to the OH + HCN case (R1b, Sec. \ref{sec:r1_gf}), the addition is the favored outcome for this pathway. However, the reaction barrier is notably smaller (by about 50 kJ $\mathrm{mol^{-1}}$) when compared to R1b. This might be imputed to the fact that the H-atom which is abstracted in R1b belongs to backbone of HCN molecule, while in R4b is part of the methyl group, hence, entailing a different kind of energetic expenditure for the process.  

\subsubsection{R4: ASW reaction pathway for OH + \ce{CH3CN}}\label{sec:r4}

\noindent \textit{Binding energy distributions and binding modes analysis}:

\begin{figure}[H]
\centering
\includegraphics[width=0.6\textwidth]{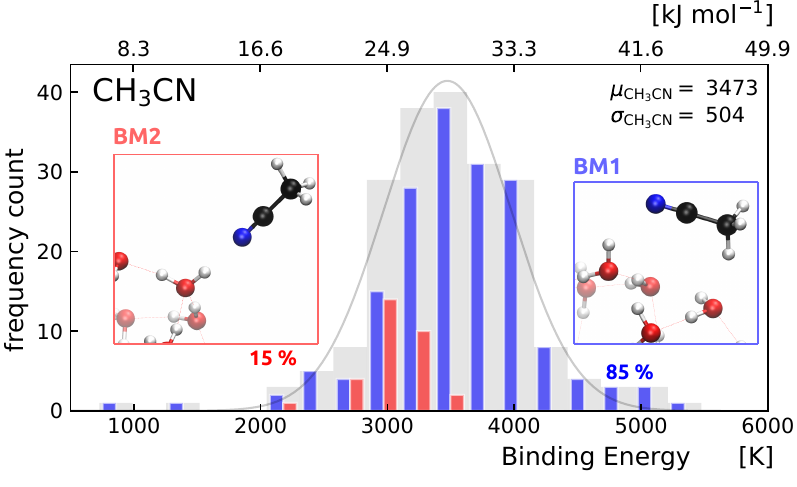}
\caption{Histogram of the BE distribution of \ce{CH3CN} computed on a set-of-clusters model 
of 22 water molecules, calculated at MPWB1K-D3BJ/def2-TZVP//HF-3c/MINIX level of theory, without including
ZPVE correction. BE values are given in K (lower scale) and kJ $\mathrm{mol^{-1}}$
(upper scale). The figure displays a composite plot: the grey histogram refers to the total data, while the structures corresponding to different binding modes (BM1-2) are colored in blue and red. 
Mean BE ($\mu$) and standard deviations of a Gaussian fit ($\sigma$) for the total distribution (grey data) are reported
in K. The insets show an
example of the different binding modes encountered.}
\label{fig:be_ch3cn}
\end{figure} 
Fig. \ref{fig:be_ch3cn} reports the BE distribution of \ce{CH3CN} adsorbed on the set of ASW clusters. The distribution contains 204 unique binding sites and its centered around 3473 K, with a standard deviation of 504 K. The $\mu$ of the distribution is shifted toward high BE values with respect to HCN by almost 650 K, while the distribution is narrower.   
Only two binding modes have been identified: one where \ce{CH3CN} interacts with the water surface via both sides (BM1, blue, analog to BM1 for HCN), and one where  only N=C is accepting a H-bond (BM2, red, analog to BM3 for HCN). We did not detect a third binding mode, where only the methyl group is involved, analog to BM2 for HCN.
BM1 structures constitute the vast majority of the binding sites (85\%). Furthermore, they represent the middle part and the entire high BE tail of the BE distribution. On the other hand, BM2 structures are mostly located around 3000 K, in correspondence of the analog  structures for HCN (BM3 sites), suggesting that, when the methyl is not taking part in the interaction, the two species present a similar adsorption behaviour.  Considering the significant difference in the percentage of structures found for the two modes, it is expected for Reaction R4(a,b) to take place in correspondence of BM1 binding sites. \\

\noindent \textit{Radical-neutral addition: formation of the imidic isomer and tautomerization to the amidic form:}

\begin{figure}[H]
\centering
\includegraphics[width=\textwidth]{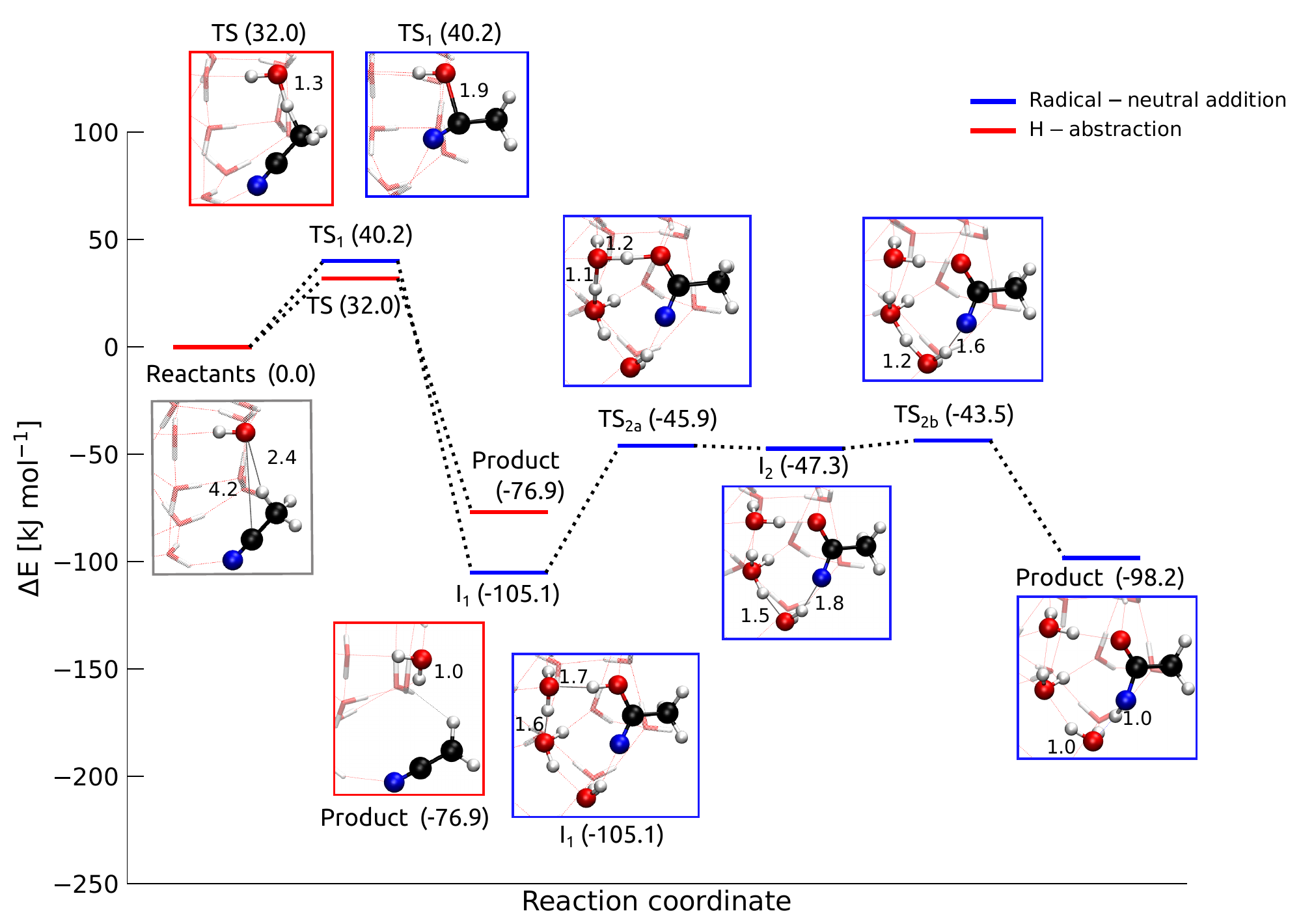}
\caption{Energy and stationary states for the ASW reaction \hg{R4}:  OH + \ce{CH3CN}, using BHandHLYP-D4/def2-SVP geometries and BMK-D3BJ/def2-TZVP
energies. The process has two possible outcomes: the addition, \hg{R4a} in blue, and the water elimination, \hg{R4b} in red.    Distances in Å. The color scheme for the atoms is red for O, black for C, blue for N and
white for H.}
\label{fig:ch3cn_oh_asw}
\end{figure} 
The results for the different reaction  
channels of the OH + \ce{CH3CN} PES  are shown in Fig. \ref{fig:ch3cn_oh_asw}.
The reactive site has been obtained by sampling a high-BE structure extracted from \ce{CH3CN} distribution. The binding site belongs to BM1 mode, as it can be seen from the H-bond interactions established by both sides of the molecule.\\

The addition reaction (Fig. \ref{fig:ch3cn_oh_asw}, blue pathway) presents a TS energy of 40.2 kJ $\mathrm{mol^{-1}}$ (\hg{Fig. \ref{fig:ch3cn_oh_asw}, \ce{TS_1}}).
The following tautomerization step from the imidic form of the radical (Fig. \ref{fig:ch3cn_oh_asw}, \hg{\ce{I_1}}) to the amidic form (\hg{Fig. \ref{fig:ch3cn_oh_asw}, Product}),
takes place  through a proton-relay involving three water molecules and follows a step-wise mechanism. The additional stationary state that we found with respect to the reaction in gas-phase,   corresponds to a dipolar intermediate (Fig. \ref{fig:ch3cn_oh_asw}, \hg{\ce{I_2}}) in which a hydronium ion is created \hg{(Fig. \ref{fig:ch3cn_oh_asw}, \hg{\ce{TS_{2a}}})}. Subsequently, in the second step of the proton-relay, 
the hydronium ion transfers the proton (Fig. \ref{fig:ch3cn_oh_asw}, \hg{\ce{TS_{2b}}}), enabling the completion of the tautomerization process. 
The stage constitute a submerged barrier with a net TS energy 
of -45.9 kJ $\mathrm{mol^{-1}}$.
Moreover, the step-wise mechanism for the tautomerization results  in an lower barrier 
(59.2 kJ $\mathrm{mol^{-1}}$), compared to the analog stage of R1a 
(68.6 kJ $\mathrm{mol^{-1}}$).   Finally, as it was the case for R1a, the imidic-acid radical isomer \hg{(Fig. \ref{fig:ch3cn_oh_asw}, \ce{I_1})} is the lowest 
energy  structure on this PES.\\ 

\noindent \textit{H-abstraction: water elimination}
The energy barrier  to the H-abstraction R4b (Fig. \ref{fig:ch3cn_oh_asw}, red pathway) is 32.0 kJ $\mathrm{mol^{-1}}$. 
Such value is slightly lower than that of the addition R4a (40.2 kJ $\mathrm{mol^{-1}}$). 
A possible reason for the lower barrier could be the particularly favorable orientation of the OH 
radical in the reactant complex on the ASW surface, which facilitates the H-abstraction reaction, 
since  the OH radical is already forming a H-bond  to the methyl H-atom. On the other hand, for
the addition (R4a) to take place, such H-bond between OH and the methyl group must be broken. 

When comparing reaction R4b on ASW to the H-abstraction on ASW in OH + HCN (R1b, Sec. \ref{sec:r1_asw}),
the H-abstraction emerges as a competitive process, as  it displays similar TS energies to the addition reaction R4a, which was not the case for R1b vs R1a.  Another difference between R1 and R4 is that R4b 
is markedly exothermic (-76.9 kJ $\mathrm{mol^{-1}}$), while R1b was endothermic by 3.1 kJ $\mathrm{mol^{-1}}$.

\subsection{Comparison between Gas-phase and ASW reaction pathways}
\hg{Table \ref{tab:bigbigtable} reports the energy barriers and reaction energies for the different pathways, carried out in gas-phase and using the ASW surface model. }\\
 \noindent \textit{\hg{R1:}} \hg{The addition of OH to HCN (R1a) on ASW sees an increase of the  TS energy with respect to the gas-phase of around 10 kJ $\mathrm{mol^{-1}}$ (Fig. \ref{fig:hcn_oh_asw}, \ref{fig:hcn_oh_gf}, \ce{TS_1}). The observed 
increase appears to stem from the enhanced interaction of the hydroxyl radical with 
the surface.  Such interaction is not present in the gas-phase and  must be overcome 
to facilitate the addition reaction, resulting in a higher \ce{TS_1} barrier. On the other 
hand, the R1a is more exothermic on the ASW surface (-138.7 vs -130.1 kJ $\mathrm{mol^{-1}}$), indicating that the interaction of Is3 radical  with the surface increases the intermediate stability (\ce{I_1} in the corresponding figures), with respect to the 
reactant complex. The tautomerization reaction  from the imidic to the amidic radical (Is3 to Is2)  on the ASW model presents a
significant  reduction compared to the gas-phase barrier (\ce{TS_2}, 68.6  vs 165.1 kJ $\mathrm{mol^{-1}}$, respectively), which is attributed to a change in mechanism, thus submerging the barrier below 
the reactant energy.}
\hg{Similarly to the addition reaction, carrying out the H-abstraction process (R1b) on ASW  
requires the breaking of the H-bond established by HCN with the surface, resulting in 
a increased barrier compared to the gas-phase (94.2 vs 82.7 kJ $\mathrm{mol^{-1}}$). On the ASW surface, the products of R1b are slighly endothermic (3.1 kJ $\mathrm{mol^{-1}}$), albeit much less than in the gas-phase.}\\  
 \noindent \textit{\hg{R2:}} \hg{In this work, the reaction has been studied solely using the ASW model, therefore, we refer to the literature\cite{rimola_can_2018} for the energy values calculated in gas-phase. 
 The reaction is the addition of CN to \ce{H2O} to form first Is4 radical (\ce{TS_1}) followed by a conformational change in the isomer configuration (\ce{TS_2}) allowing for the tautomerization (\ce{TS_3}) from Is4 to Is1 carbamoyl radical. 
 All the reaction barrier are notably lower on the ASW compared to the gas-phase, by around 100 kJ $\mathrm{mol^{-1}}$ for the first two steps, and even more (184.6 kJ $\mathrm{mol^{-1}}$) for \ce{TS_3}. The third step is also the one that differ the most in terms of reaction mechanism, as it is concerted in the gas-phase, while it takes place in a step-wise fashion on ASW. 
 The change can be imputed to the fact that in the latter, the tautomerization takes place establishing a proton-relay that involves three water molecules from the ASW, bridging the proton from one side of the radical to the other. 
 That feature increases the complexity of the process, since the H-bond network of the ice molecules has a important and specific effect of each step of the proton transfer process. Both the reactions in gas-phase and on ASW have a marked exothermic character. 
 In summary, the major difference between R2 carried out in the gas-phase, is constituted by the fact that, on ASW, the barriers for \ce{TS_{2,3}} are lowered to such extent that  become submerged with respect to the reactant energy, making \ce{TS_1} the only barrier to be overcome for the process to take place.}\\
 \noindent \textit{\hg{R3:}} \hg{The addition of \ce{NH2} to CO is the process that presents the smallest variations  between the gas-phase and the reaction on ASW. 
 In fact, there is no change in the mechanism, and the TS barriers are comparable (12.6 vs 10.2 kJ $\mathrm{mol^{-1}}$ for ASW and gas-phase, respectively) as well as the reaction exothermicity (-110.8 vs -120.0 kJ $\mathrm{mol^{-1}}$). This is due to the fact that, in the ASW reactive site,  only one of the reactive fragments (\ce{NH2}) is interacting with the ice, hence, the TS geometry is  mostly unchanged with respect to the gas-phase.}\\
\noindent \textit{\hg{R4:}} \hg{ The reaction between OH and \ce{CH3CN} presents to possible outcomes. The addition pathway (R4a) on ASW presents a  noticeably  higher barrier (almost 20 kJ $\mathrm{mol^{-1}}$) than the gas-phase reaction. 
Such a increase in the TS energy stems from the orientation of the OH radical in the reactive complex, 
which is aligned towards the methyl group and thus more impeded in forming
the C--O bond. 
Regarding the second step of the reaction (\ce{TS_2}), the tautomerization of the imidic isomer to the amidic form of the radical,  the mechanism  changes from concerted in the gas-phase  to step-wise on ASW, where the proton transfer takes place via a proton relay involving three water molecules, similarly to what we found for R2.  
The total barrier for this process (59.2  kJ $\mathrm{mol^{-1}}$) is also much lower than that of the concerted reaction in the gas-phase, by about 100 kJ $\mathrm{mol^{-1}}$. Moreover, we found \ce{TS_2} for such pathway to become a submerged barrier,  in passing from the gas-phase to ASW, as it was the case for R2. 
On the other hand, the barrier for the H-abstraction process (R4b), on ASW, is very similar to that in gas-phase (32.0 vs 32.9 kJ $\mathrm{mol^{-1}}$): the aforementioned orientation of the OH radical in the reactive complex, which  minimizes the distance with the methyl group, results to be particularly favorable for this specific outcome.}

\section{Astrophysical implications}

Based on the results presented in this paper, the
addition path OH + HCN to yield the N-radical form of the 
carbamoyl (Is2), represents an alternative towards amides formation,  
considering a net reaction barrier of 26.7 kJ/mol (3200 K). Although, the energy barrier 
makes it unlikely for it to take place 
in cold dark molecular clouds, the reaction might
play a role in later stages of star forming regions.
Hence, in the context of this study, a high-BE binding site for HCN makes it more 
likely to encounter diffusing OH radicals during the warm up phase, before 
evaporation of the ASW layer. 

The selected HCN binding site presented a BE of 4500 K, that is larger  than both the 
OH addition barrier estimated in this work,  and the average BE of ASW of 4200 K \cite{Tinacci_2023}. 

Furthermore, we found that the resulting 
N-radical form of the carbamoyl (Is2), product of OH addition, has an exceptional affinity to the ice surface. Estimation of the BE of such radical on a set of \ce{(H2O)_{1--5}} clusters revealed a average BE of 9931 K, the highest among the carbamoyl isomers considered in this work. 
Our result suggests that such radical Is2, being more strongly bound to ASW, with respect to carbamoyl (Is1), might survive hydrogenation processes leading to direct generation of formamide (Reaction \ref{eq:1}). Such reaction is object of debate\cite{noble_hydrogenation_2015,haupa_hydrogen_2019}, given the competitive channel for the production HNCO:

\begin{equation}
    \ce{NH2CO} + \ce{H} \longrightarrow \ce{HNCO} + \ce{H2}
\end{equation}

Moreover, Is2 isomer is  linked to another important interstellar amides:  
radical recombination of Is2  with a \ce{CH3} radical leads to 
formation of N-methylformamide (Reaction \ref{eq:10}).\\

Taking into account the alternative reactive channel involving   
H-abstraction from HCN, leading to water elimination,  
revealed a considerably higher barrier 
with respect to OH-addition reaction, by 66.7 kJ/mol.  
Nevertheless,   the reaction 
might still be viable under interstellar conditions, due to the increase in the  
reaction rate  when considering  quantum tunneling effects on  the abstracted hydrogen.

The study of the addition of CN radical to a water molecule strongly bound to the 
ice surface, leading to the formation of 
H(N=C)OH 
isomer of carbamoyl (Is4), subsequently converted to Is1, displays a high reaction barrier of 70.6 kJ/mol 
(8400 K), which makes it overall an unlikely surface route.

Finally the \ce{NH2 + CO} pathway  displays the lowest barrier of 
all reactions considered in this work for the generation of carbamoyl isomers. 
The barrier  on ASW is 12.6 kJ/mol (1500 K), which is 1000 K lower than the 
barriers used in astrochemical models that include this reaction\cite{garrod_three-phase_2013}. Our result suggests the need to reconsider the treatment of this process in kinetic models, in order to properly account for the surface catalytic effect provided by the ASW ice. \\
Finally, the analysis of the possible formation  pathway for N-radical acetamide  precursor (CH3(C=O)NH), via the addition reaction of OH to  \ce{CH3CN}, resulted to be less viable than the analog reaction between the radical and HCN.
The result is imputed to the higher reaction barrier for the process on the ASW surface, due to the influence of H-bond network on OH orientation with respect to \ce{CH3CN} molecule,     
along with the  higher 
affinity of the OH towards the methyl hydrogen of \ce{CH3CN}, which favors the H-abstraction competitive outcome. Therefore, 
the most probable result of the reaction  
 will be the formation 
of \ce{CH2CN} species, rather than the acetamide precursor radical, more so, considering the  possibility of tunneling effects.
\hg{Our results suggest alternative reaction pathways for amides in the interstellar medium. To improve the detection of these amide species with ALMA, it is essential to conduct more detailed studies on their infrared spectrum and millimeter rotational features in the gas phase. Such studies are particularly important in regions where the warmth from newborn stars releases these prebiotic precursors into the gas phase. Moreover, the high average  binding energies of the resulting amide radicals and the absence of amide observations in cold prestellar cores, suggest that the amides and its precursors would remain frozen out onto the grain surface until
the warming up phase of  the protostellar core, where they have been detected. Future surveys utilizing the James Webb Space Telescope (JWST), with its exceptional resolution,  could facilitate the integration of experimental data with observational findings.}

\section{Conclusions}

In this work, we explored different radical-neutral formation pathways towards simple
interstellar amides,   
on a model ASW surface, through the formation of carbamoyl
\ce{NH2CO} radical isomers.  We computed the relative  energy of four carbamoyl isomers in the gas-phase  and obtained a range  of binding energies for the  isomers adsorbed
on small water clusters 
(\ce{H2O})$_{(1-5)}$. We evaluated,  by means of DFT, the binding energy distribution of the 
reactant molecules \ce{HCN} and \ce{CH3CN}  on a ASW set-of-cluster model consisting 
of 22-water molecules each,  and analyzed the different binding modes 
to select a high binding energy binding mode to study the OH addition reactions.
The main conclusions that can be drawn from this study are the following:

\begin{enumerate}
    \item In the gas-phase, the lowest energy isomer of the \ce{NH2CO} radical 
     family is the C-radical carbamoyl (isomer Is1, Fig.\ref{fig:isomers}),  at the UCCSD(T)-F12/cc-PVDZ-F12 level 
     of theory
     \item The isomer with the highest average BE 
     is the N-radical  form (\ce{NH(C=O)H}, isomer Is2, Fig.\ref{fig:isomers}). Surprisingly, carbamoyl (\ce{NH2CO}, Is1),
     displays the lowest binding energy. This 
     suggests a markedly different stability profile for both radicals on an 
     ice surface, when compared to the gas-phase.
     More detailed analysis of the adsorption motives of such species on the ASW surface is the object of a forthcoming study.
     \item The OH + HCN addition pathway, resulting in the imidic form of the carbamoyl radical, followed by the tautomerization to the N-radical form (Is3 to Is2),   
     appears to be a feasible formation pathway for formamide and N-methylformamide.  In fact, the barrier for OH addition is lower than the binding energy of  HCN in the reactive site, such that 
     reactive encounters with diffusing OH radicals might take place under interstellar condition akin to those of star-forming regions. 
     \item The reaction of CN radical with a water molecule belonging to the ASW surface, 
     displays a high reaction barrier, making this pathway unlikely 
     to take place even in warmer star-forming region.
     \item The addition reaction of \ce{NH2} to CO is the energetically 
     most favorable formation route for the generation of carbamoyl \ce{NH2CO}, 
     with a modest activation barrier of  12.6 kJ $\mathrm{mol^{-1}}$ (1500 K). This barrier is 1000 K
     lower than the barrier currently used in astrochemical models, suggesting
     the need to update this value in future models.
     \item The formation of acetamide precursor through the addition reaction of OH-radicals 
     to \ce{CH3CN}, seems unlikely to take place on ASW surface. We found the barrier to the formation of the imidic radical intermediate (CH3C(OH)N) to be significantly higher than the corresponding  gas-phase barrier,  resulting in a anti-catalytic
     behaviour of the ASW surface. Moreover, the lower TS energy displayed by the competitive H-abstraction reaction channel, makes the formation of the radical species  (\ce{CH2CN}) the most plausible outcome.
\end{enumerate}

These findings suggest potential alternative pathways for the formation of amides through the generation of various isomers of carbamoyl (\ce{NH2CO}) radical.
However, examining the stability of the different isomers with respect to
hydrogenation and methylation on ASW will be  pivotal to 
shed light on the role of such radical precursors in the chemistry of interstellar amides.

\begin{acknowledgement}

SVG thanks VRID research grant 2022000507INV for financing this project. GMB gratefully acknowledges support from ANID Beca de Doctorado Nacional 21200180. GSV thanks the Universidad de Concepción for the Beca de Excelencia Académica scholarship. NR thanks FONDECYT POSTDOCTORADO grant 3230221 for financial support. 

\end{acknowledgement}

\begin{suppinfo}

Supporting Information: XYZ coordinates of all the stationary points
reported in this work.

\end{suppinfo}

\providecommand{\latin}[1]{#1}
\makeatletter
\providecommand{\doi}
  {\begingroup\let\do\@makeother\dospecials
  \catcode`\{=1 \catcode`\}=2 \doi@aux}
\providecommand{\doi@aux}[1]{\endgroup\texttt{#1}}
\makeatother
\providecommand*\mcitethebibliography{\thebibliography}
\csname @ifundefined\endcsname{endmcitethebibliography}
  {\let\endmcitethebibliography\endthebibliography}{}


\begin{mcitethebibliography}{52}
\providecommand*\natexlab[1]{#1}
\providecommand*\mciteSetBstSublistMode[1]{}
\providecommand*\mciteSetBstMaxWidthForm[2]{}
\providecommand*\mciteBstWouldAddEndPuncttrue
  {\def\EndOfBibitem{\unskip.}}
\providecommand*\mciteBstWouldAddEndPunctfalse
  {\let\EndOfBibitem\relax}
\providecommand*\mciteSetBstMidEndSepPunct[3]{}
\providecommand*\mciteSetBstSublistLabelBeginEnd[3]{}
\providecommand*\EndOfBibitem{}
\mciteSetBstSublistMode{f}
\mciteSetBstMaxWidthForm{subitem}{(\alph{mcitesubitemcount})}
\mciteSetBstSublistLabelBeginEnd
  {\mcitemaxwidthsubitemform\space}
  {\relax}
  {\relax}

\bibitem[McClure \latin{et~al.}(2023)McClure, Rocha, Pontoppidan, Crouzet, Chu,
  Dartois, Lamberts, Noble, Pendleton, Perotti, Qasim, Rachid, Smith, Sun,
  Beck, Boogert, Brown, Caselli, Charnley, Cuppen, Dickinson, Drozdovskaya,
  Egami, Erkal, Fraser, Garrod, Harsono, Ioppolo, JimÃ©nez-Serra, Jin,
  JÃžrgensen, Kristensen, Lis, McCoustra, McGuire, Melnick, Ãberg, Palumbo,
  Shimonishi, Sturm, van Dishoeck, and Linnartz]{mcclure_ice_2023}
McClure,~M.~K. \latin{et~al.}  An {Ice} {Age} {JWST} inventory of dense
  molecular cloud ices. \emph{Nature Astronomy} \textbf{2023}, \emph{7},
  431--443, Number: 4 Publisher: Nature Publishing Group\relax
\mciteBstWouldAddEndPuncttrue
\mciteSetBstMidEndSepPunct{\mcitedefaultmidpunct}
{\mcitedefaultendpunct}{\mcitedefaultseppunct}\relax
\EndOfBibitem
\bibitem[Portugal \latin{et~al.}(2014)Portugal, Pilling, Boduch, Rothard, and
  Andrade]{portugal_radiolysis_2014}
Portugal,~W.; Pilling,~S.; Boduch,~P.; Rothard,~H.; Andrade,~D. P.~P.
  Radiolysis of amino acids by heavy and energetic cosmic ray analogues in
  simulated space environments: Î±-glycine zwitterion form. \emph{Monthly
  Notices of the Royal Astronomical Society} \textbf{2014}, \emph{441},
  3209--3225\relax
\mciteBstWouldAddEndPuncttrue
\mciteSetBstMidEndSepPunct{\mcitedefaultmidpunct}
{\mcitedefaultendpunct}{\mcitedefaultseppunct}\relax
\EndOfBibitem
\bibitem[LÃ³pez-Sepulcre \latin{et~al.}(2019)LÃ³pez-Sepulcre, Balucani,
  Ceccarelli, Codella, Dulieu, and TheulÃ©]{lopez-sepulcre_interstellar_2019}
LÃ³pez-Sepulcre,~A.; Balucani,~N.; Ceccarelli,~C.; Codella,~C.; Dulieu,~F.;
  TheulÃ©,~P. Interstellar {Formamide} ({NH2CHO}), a {Key} {Prebiotic}
  {Precursor}. \emph{ACS Earth and Space Chemistry} \textbf{2019}, \emph{3},
  2122--2137, Publisher: American Chemical Society\relax
\mciteBstWouldAddEndPuncttrue
\mciteSetBstMidEndSepPunct{\mcitedefaultmidpunct}
{\mcitedefaultendpunct}{\mcitedefaultseppunct}\relax
\EndOfBibitem
\bibitem[Colzi \latin{et~al.}(2021)Colzi, Rivilla, BeltrÃ¡n, JimÃ©nez-Serra,
  Mininni, Melosso, Cesaroni, Fontani, Lorenzani, SÃ¡nchez-Monge, Viti,
  Schilke, Testi, Alonso, and KolesnikovÃ¡]{colzi_guapos_2021}
Colzi,~L.; Rivilla,~V.~M.; BeltrÃ¡n,~M.~T.; JimÃ©nez-Serra,~I.; Mininni,~C.;
  Melosso,~M.; Cesaroni,~R.; Fontani,~F.; Lorenzani,~A.; SÃ¡nchez-Monge,~A.;
  Viti,~S.; Schilke,~P.; Testi,~L.; Alonso,~E.~R.; KolesnikovÃ¡,~L. The
  {GUAPOS} project - {II}. {A} comprehensive study of peptide-like bond
  molecules. \emph{Astronomy \& Astrophysics} \textbf{2021}, \emph{653}, A129,
  Publisher: EDP Sciences\relax
\mciteBstWouldAddEndPuncttrue
\mciteSetBstMidEndSepPunct{\mcitedefaultmidpunct}
{\mcitedefaultendpunct}{\mcitedefaultseppunct}\relax
\EndOfBibitem
\bibitem[Ligterink \latin{et~al.}(2020)Ligterink, El-Abd, Brogan, Hunter,
  Remijan, Garrod, and McGuire]{ligterink_family_2020}
Ligterink,~N. F.~W.; El-Abd,~S.~J.; Brogan,~C.~L.; Hunter,~T.~R.;
  Remijan,~A.~J.; Garrod,~R.~T.; McGuire,~B.~M. The {Family} of {Amide}
  {Molecules} toward {NGC} {6334I}. \emph{The Astrophysical Journal}
  \textbf{2020}, \emph{901}, 37, Publisher: American Astronomical Society\relax
\mciteBstWouldAddEndPuncttrue
\mciteSetBstMidEndSepPunct{\mcitedefaultmidpunct}
{\mcitedefaultendpunct}{\mcitedefaultseppunct}\relax
\EndOfBibitem
\bibitem[Ligterink \latin{et~al.}(2022)Ligterink, Ahmadi, Luitel, Coutens,
  Calcutt, Tychoniec, Linnartz, JÃžrgensen, Garrod, and
  Bouwman]{ligterink_prebiotic_2022}
Ligterink,~N. F.~W.; Ahmadi,~A.; Luitel,~B.; Coutens,~A.; Calcutt,~H.;
  Tychoniec,~Å.; Linnartz,~H.; JÃžrgensen,~J.~K.; Garrod,~R.~T.; Bouwman,~J.
  The prebiotic molecular inventory of {Serpens} {SMM1}: {II}. {The} building
  blocks of peptide chains. 2022; \url{http://arxiv.org/abs/2202.09640},
  arXiv:2202.09640 [astro-ph]\relax
\mciteBstWouldAddEndPuncttrue
\mciteSetBstMidEndSepPunct{\mcitedefaultmidpunct}
{\mcitedefaultendpunct}{\mcitedefaultseppunct}\relax
\EndOfBibitem
\bibitem[Zeng \latin{et~al.}(2023)Zeng, Rivilla, JimÃ©nez-Serra, Colzi,
  MartÃ­n-Pintado, Tercero, deÂ Vicente, MartÃ­n, and
  Requena-Torres]{zeng_amides_2023}
Zeng,~S.; Rivilla,~V.~M.; JimÃ©nez-Serra,~I.; Colzi,~L.; MartÃ­n-Pintado,~J.;
  Tercero,~B.; deÂ Vicente,~P.; MartÃ­n,~S.; Requena-Torres,~M.~A. Amides
  inventory towards the {G}+0.693â0.027 molecular cloud. \emph{Monthly
  Notices of the Royal Astronomical Society} \textbf{2023}, \emph{523},
  1448--1463\relax
\mciteBstWouldAddEndPuncttrue
\mciteSetBstMidEndSepPunct{\mcitedefaultmidpunct}
{\mcitedefaultendpunct}{\mcitedefaultseppunct}\relax
\EndOfBibitem
\bibitem[Ligterink \latin{et~al.}(2022)Ligterink, Ahmadi, Luitel, Coutens,
  Calcutt, Tychoniec, Linnartz, JÃžrgensen, Garrod, and
  Bouwman]{ligterink_prebiotic_2022-2}
Ligterink,~N. F.~W.; Ahmadi,~A.; Luitel,~B.; Coutens,~A.; Calcutt,~H.;
  Tychoniec,~Å.; Linnartz,~H.; JÃžrgensen,~J.~K.; Garrod,~R.~T.; Bouwman,~J.
  The {Prebiotic} {Molecular} {Inventory} of {Serpens} {SMM1}: {II}. {The}
  {Building} {Blocks} of {Peptide} {Chains}. \emph{ACS Earth and Space
  Chemistry} \textbf{2022}, \emph{6}, 455--467, Publisher: American Chemical
  Society\relax
\mciteBstWouldAddEndPuncttrue
\mciteSetBstMidEndSepPunct{\mcitedefaultmidpunct}
{\mcitedefaultendpunct}{\mcitedefaultseppunct}\relax
\EndOfBibitem
\bibitem[Sanz-Novo \latin{et~al.}(2022)Sanz-Novo, Belloche, Rivilla, Garrod,
  Alonso, Redondo, Barrientos, KolesnikovÃ¡, Valle, RodrÃ­guez-Almeida,
  Jimenez-Serra, MartÃ­n-Pintado, MÃŒller, and Menten]{sanz-novo_toward_2022}
Sanz-Novo,~M.; Belloche,~A.; Rivilla,~V.~M.; Garrod,~R.~T.; Alonso,~J.~L.;
  Redondo,~P.; Barrientos,~C.; KolesnikovÃ¡,~L.; Valle,~J.~C.;
  RodrÃ­guez-Almeida,~L.; Jimenez-Serra,~I.; MartÃ­n-Pintado,~J.; MÃŒller,~H.
  S.~P.; Menten,~K.~M. Toward the limits of complexity of interstellar
  chemistry: {Rotational} spectroscopy and astronomical search for n- and
  i-butanal. \emph{Astronomy \& Astrophysics} \textbf{2022}, \emph{666}, A114,
  Publisher: EDP Sciences\relax
\mciteBstWouldAddEndPuncttrue
\mciteSetBstMidEndSepPunct{\mcitedefaultmidpunct}
{\mcitedefaultendpunct}{\mcitedefaultseppunct}\relax
\EndOfBibitem
\bibitem[Bisschop \latin{et~al.}(2007)Bisschop, JÃžrgensen, van Dishoeck, and
  de~Wachter]{bisschop_testing_2007}
Bisschop,~S.~E.; JÃžrgensen,~J.~K.; van Dishoeck,~E.~F.; de~Wachter,~E. B.~M.
  Testing grain-surface chemistry in massive hot-core regions. \emph{Astronomy
  \& Astrophysics} \textbf{2007}, \emph{465}, 913--929\relax
\mciteBstWouldAddEndPuncttrue
\mciteSetBstMidEndSepPunct{\mcitedefaultmidpunct}
{\mcitedefaultendpunct}{\mcitedefaultseppunct}\relax
\EndOfBibitem
\bibitem[LÃ³pez-Sepulcre \latin{et~al.}(2015)LÃ³pez-Sepulcre, Jaber, Mendoza,
  Lefloch, Ceccarelli, Vastel, Bachiller, Cernicharo, Codella, Kahane, Kama,
  and Tafalla]{lopez-sepulcre_shedding_2015}
LÃ³pez-Sepulcre,~A.; Jaber,~A.~A.; Mendoza,~E.; Lefloch,~B.; Ceccarelli,~C.;
  Vastel,~C.; Bachiller,~R.; Cernicharo,~J.; Codella,~C.; Kahane,~C.; Kama,~M.;
  Tafalla,~M. Shedding light on the formation of the pre-biotic molecule
  formamide with {ASAI}. \emph{Monthly Notices of the Royal Astronomical
  Society} \textbf{2015}, \emph{449}, 2438--2458, arXiv:1502.05762
  [astro-ph]\relax
\mciteBstWouldAddEndPuncttrue
\mciteSetBstMidEndSepPunct{\mcitedefaultmidpunct}
{\mcitedefaultendpunct}{\mcitedefaultseppunct}\relax
\EndOfBibitem
\bibitem[Hubbard \latin{et~al.}(1975)Hubbard, Voecks, Hobby, Ferris, Williams,
  and Nicodem]{hubbard_ultraviolet-gas_1975}
Hubbard,~J.~S.; Voecks,~G.~E.; Hobby,~G.~L.; Ferris,~J.~P.; Williams,~E.~A.;
  Nicodem,~D.~E. Ultraviolet-gas phase and -photocatalytic synthesis from {CO}
  and {NH3}. \emph{Journal of Molecular Evolution} \textbf{1975}, \emph{5},
  223--241\relax
\mciteBstWouldAddEndPuncttrue
\mciteSetBstMidEndSepPunct{\mcitedefaultmidpunct}
{\mcitedefaultendpunct}{\mcitedefaultseppunct}\relax
\EndOfBibitem
\bibitem[Agarwal \latin{et~al.}(1985)Agarwal, Schutte, Greenberg, Ferris,
  Briggs, Connor, Van~de Bult, and Baas]{agarwal_photochemical_1985}
Agarwal,~V.~K.; Schutte,~W.; Greenberg,~J.~M.; Ferris,~J.~P.; Briggs,~R.;
  Connor,~S.; Van~de Bult,~C. P. E.~M.; Baas,~F. Photochemical reactions in
  interstellar grains photolysis of co, {NH3}, and {H2O}. \emph{Origins of life
  and evolution of the biosphere} \textbf{1985}, \emph{16}, 21--40\relax
\mciteBstWouldAddEndPuncttrue
\mciteSetBstMidEndSepPunct{\mcitedefaultmidpunct}
{\mcitedefaultendpunct}{\mcitedefaultseppunct}\relax
\EndOfBibitem
\bibitem[Ligterink \latin{et~al.}(2018)Ligterink, TerwisschaÂ vanÂ Scheltinga,
  Taquet, JÃžrgensen, Cazaux, vanÂ Dishoeck, and
  Linnartz]{ligterink_formation_2018}
Ligterink,~N. F.~W.; TerwisschaÂ vanÂ Scheltinga,~J.; Taquet,~V.;
  JÃžrgensen,~J.~K.; Cazaux,~S.; vanÂ Dishoeck,~E.~F.; Linnartz,~H. The
  formation of peptide-like molecules on interstellar dust grains.
  \emph{Monthly Notices of the Royal Astronomical Society} \textbf{2018},
  \emph{480}, 3628--3643\relax
\mciteBstWouldAddEndPuncttrue
\mciteSetBstMidEndSepPunct{\mcitedefaultmidpunct}
{\mcitedefaultendpunct}{\mcitedefaultseppunct}\relax
\EndOfBibitem
\bibitem[Chuang \latin{et~al.}(2022)Chuang, JÃ€ger, Krasnokutski, Fulvio, and
  Henning]{chuang_formation_2022}
Chuang,~K.-J.; JÃ€ger,~C.; Krasnokutski,~S.~A.; Fulvio,~D.; Henning,~T.
  Formation of the {Simplest} {Amide} in {Molecular} {Clouds}: {Formamide}
  ({NH} $_{\textrm{2}}$ {CHO}) and {Its} {Derivatives} in {H} $_{\textrm{2}}$
  {O}-rich and {CO}-rich {Interstellar} {Ice} {Analogs} upon {VUV}
  {Irradiation}. \emph{The Astrophysical Journal} \textbf{2022}, \emph{933},
  107\relax
\mciteBstWouldAddEndPuncttrue
\mciteSetBstMidEndSepPunct{\mcitedefaultmidpunct}
{\mcitedefaultendpunct}{\mcitedefaultseppunct}\relax
\EndOfBibitem
\bibitem[Noble \latin{et~al.}(2015)Noble, Theule, Congiu, Dulieu, Bonnin,
  Bassas, Duvernay, Danger, and Chiavassa]{noble_hydrogenation_2015}
Noble,~J.~A.; Theule,~P.; Congiu,~E.; Dulieu,~F.; Bonnin,~M.; Bassas,~A.;
  Duvernay,~F.; Danger,~G.; Chiavassa,~T. Hydrogenation at low temperatures
  does not always lead to saturation: the case of {HNCO}. \emph{Astronomy \&
  Astrophysics} \textbf{2015}, \emph{576}, A91\relax
\mciteBstWouldAddEndPuncttrue
\mciteSetBstMidEndSepPunct{\mcitedefaultmidpunct}
{\mcitedefaultendpunct}{\mcitedefaultseppunct}\relax
\EndOfBibitem
\bibitem[Haupa \latin{et~al.}(2019)Haupa, Tarczay, and
  Lee]{haupa_hydrogen_2019}
Haupa,~K.~A.; Tarczay,~G.; Lee,~Y.-P. Hydrogen {Abstraction}/{Addition}
  {Tunneling} {Reactions} {Elucidate} the {Interstellar} {H} $_{\textrm{2}}$
  {NCHO}/{HNCO} {Ratio} and {H} $_{\textrm{2}}$ {Formation}. \emph{Journal of
  the American Chemical Society} \textbf{2019}, \emph{141}, 11614--11620\relax
\mciteBstWouldAddEndPuncttrue
\mciteSetBstMidEndSepPunct{\mcitedefaultmidpunct}
{\mcitedefaultendpunct}{\mcitedefaultseppunct}\relax
\EndOfBibitem
\bibitem[Slate \latin{et~al.}(2020)Slate, Barker, Euesden, Revels, and
  Meijer]{slate_computational_2020}
Slate,~E. C.~S.; Barker,~R.; Euesden,~R.~T.; Revels,~M.~R.; Meijer,~A. J. H.~M.
  Computational studies into urea formation in the interstellar medium.
  \emph{Monthly Notices of the Royal Astronomical Society} \textbf{2020},
  \emph{497}, 5413--5420\relax
\mciteBstWouldAddEndPuncttrue
\mciteSetBstMidEndSepPunct{\mcitedefaultmidpunct}
{\mcitedefaultendpunct}{\mcitedefaultseppunct}\relax
\EndOfBibitem
\bibitem[Belloche \latin{et~al.}(2017)Belloche, Meshcheryakov, Garrod,
  Ilyushin, Alekseev, Motiyenko, MargulÃšs, MÃŒller, and
  Menten]{belloche_rotational_2017}
Belloche,~A.; Meshcheryakov,~A.~A.; Garrod,~R.~T.; Ilyushin,~V.~V.;
  Alekseev,~E.~A.; Motiyenko,~R.~A.; MargulÃšs,~L.; MÃŒller,~H. S.~P.;
  Menten,~K.~M. Rotational spectroscopy, tentative interstellar detection, and
  chemical modeling of {N}-methylformamide. \emph{Astronomy \& Astrophysics}
  \textbf{2017}, \emph{601}, A49, Publisher: EDP Sciences\relax
\mciteBstWouldAddEndPuncttrue
\mciteSetBstMidEndSepPunct{\mcitedefaultmidpunct}
{\mcitedefaultendpunct}{\mcitedefaultseppunct}\relax
\EndOfBibitem
\bibitem[Belloche \latin{et~al.}(2019)Belloche, Garrod, MÃŒller, Menten,
  Medvedev, Thomas, and Kisiel]{belloche_re-exploring_2019}
Belloche,~A.; Garrod,~R.~T.; MÃŒller,~H. S.~P.; Menten,~K.~M.; Medvedev,~I.;
  Thomas,~J.; Kisiel,~Z. Re-exploring {Molecular} {Complexity} with {ALMA}
  ({ReMoCA}): interstellar detection of urea. \emph{Astronomy \& Astrophysics}
  \textbf{2019}, \emph{628}, A10, Publisher: EDP Sciences\relax
\mciteBstWouldAddEndPuncttrue
\mciteSetBstMidEndSepPunct{\mcitedefaultmidpunct}
{\mcitedefaultendpunct}{\mcitedefaultseppunct}\relax
\EndOfBibitem
\bibitem[Garrod \latin{et~al.}(2022)Garrod, Jin, Matis, Jones, Willis, and
  Herbst]{garrod_formation_2022}
Garrod,~R.~T.; Jin,~M.; Matis,~K.~A.; Jones,~D.; Willis,~E.~R.; Herbst,~E.
  Formation of {Complex} {Organic} {Molecules} in {Hot} {Molecular} {Cores}
  through {Nondiffusive} {Grain}-surface and {Ice}-mantle {Chemistry}.
  \emph{The Astrophysical Journal Supplement Series} \textbf{2022}, \emph{259},
  1, Publisher: The American Astronomical Society\relax
\mciteBstWouldAddEndPuncttrue
\mciteSetBstMidEndSepPunct{\mcitedefaultmidpunct}
{\mcitedefaultendpunct}{\mcitedefaultseppunct}\relax
\EndOfBibitem
\bibitem[Hudson and Moore(2000)Hudson, and Moore]{hudson_new_2000}
Hudson,~R.~L.; Moore,~M.~H. New experiments and interpretations concerning the
  ``{XCN}'' band in interstellar ice analogues. \emph{Astronomy and
  Astrophysics} \textbf{2000}, \emph{357}, 787--792, ADS Bibcode:
  2000A\&A...357..787H\relax
\mciteBstWouldAddEndPuncttrue
\mciteSetBstMidEndSepPunct{\mcitedefaultmidpunct}
{\mcitedefaultendpunct}{\mcitedefaultseppunct}\relax
\EndOfBibitem
\bibitem[BredehÃ¶ft \latin{et~al.}(2017)BredehÃ¶ft, BÃ¶hler, Schmidt, Borrmann,
  and Swiderek]{bredehoft_electron-induced_2017}
BredehÃ¶ft,~J.~H.; BÃ¶hler,~E.; Schmidt,~F.; Borrmann,~T.; Swiderek,~P.
  Electron-{Induced} {Synthesis} of {Formamide} in {Condensed} {Mixtures} of
  {Carbon} {Monoxide} and {Ammonia}. \emph{ACS Earth and Space Chemistry}
  \textbf{2017}, \emph{1}, 50--59\relax
\mciteBstWouldAddEndPuncttrue
\mciteSetBstMidEndSepPunct{\mcitedefaultmidpunct}
{\mcitedefaultendpunct}{\mcitedefaultseppunct}\relax
\EndOfBibitem
\bibitem[Gerakines \latin{et~al.}(2004)Gerakines, Moore, and
  Hudson]{gerakines_ultraviolet_2004}
Gerakines,~P.; Moore,~M.; Hudson,~R. Ultraviolet photolysis and proton
  irradiation of astrophysical ice analogs containing hydrogen cyanide.
  \emph{Icarus} \textbf{2004}, \emph{170}, 202--213\relax
\mciteBstWouldAddEndPuncttrue
\mciteSetBstMidEndSepPunct{\mcitedefaultmidpunct}
{\mcitedefaultendpunct}{\mcitedefaultseppunct}\relax
\EndOfBibitem
\bibitem[Rimola \latin{et~al.}(2018)Rimola, Skouteris, Balucani, Ceccarelli,
  Enrique-Romero, Taquet, and Ugliengo]{rimola_can_2018}
Rimola,~A.; Skouteris,~D.; Balucani,~N.; Ceccarelli,~C.; Enrique-Romero,~J.;
  Taquet,~V.; Ugliengo,~P. Can {Formamide} {Be} {Formed} on {Interstellar}
  {Ice}? {An} {Atomistic} {Perspective}. \emph{ACS Earth and Space Chemistry}
  \textbf{2018}, \emph{2}, 720--734, Publisher: American Chemical Society\relax
\mciteBstWouldAddEndPuncttrue
\mciteSetBstMidEndSepPunct{\mcitedefaultmidpunct}
{\mcitedefaultendpunct}{\mcitedefaultseppunct}\relax
\EndOfBibitem
\bibitem[Hama and Watanabe(2013)Hama, and Watanabe]{hama_surface_2013}
Hama,~T.; Watanabe,~N. Surface {Processes} on {Interstellar} {Amorphous}
  {Solid} {Water}: {Adsorption}, {Diffusion}, {Tunneling} {Reactions}, and
  {Nuclear}-{Spin} {Conversion}. \emph{Chemical Reviews} \textbf{2013},
  \emph{113}, 8783--8839, Publisher: American Chemical Society\relax
\mciteBstWouldAddEndPuncttrue
\mciteSetBstMidEndSepPunct{\mcitedefaultmidpunct}
{\mcitedefaultendpunct}{\mcitedefaultseppunct}\relax
\EndOfBibitem
\bibitem[He \latin{et~al.}(2011)He, Frank, and Vidali]{he_interaction_2011}
He,~J.; Frank,~P.; Vidali,~G. Interaction of hydrogen with surfaces of
  silicates: single crystal vs. amorphous. \emph{Physical Chemistry Chemical
  Physics} \textbf{2011}, \emph{13}, 15803--15809, Publisher: The Royal Society
  of Chemistry\relax
\mciteBstWouldAddEndPuncttrue
\mciteSetBstMidEndSepPunct{\mcitedefaultmidpunct}
{\mcitedefaultendpunct}{\mcitedefaultseppunct}\relax
\EndOfBibitem
\bibitem[Noble \latin{et~al.}(2012)Noble, Congiu, Dulieu, and
  Fraser]{noble_thermal_2012}
Noble,~J.~A.; Congiu,~E.; Dulieu,~F.; Fraser,~H.~J. Thermal desorption
  characteristics of {CO}, {O2} and {CO2} on non-porous water, crystalline
  water and silicate surfaces at submonolayer and multilayer coverages.
  \emph{Monthly Notices of the Royal Astronomical Society} \textbf{2012},
  \emph{421}, 768--779, Publisher: Oxford Academic\relax
\mciteBstWouldAddEndPuncttrue
\mciteSetBstMidEndSepPunct{\mcitedefaultmidpunct}
{\mcitedefaultendpunct}{\mcitedefaultseppunct}\relax
\EndOfBibitem
\bibitem[Grassi \latin{et~al.}(2020)Grassi, Bovino, Caselli, Bovolenta,
  Vogt-Geisse, and Ercolano]{grassi_novel_2020}
Grassi,~T.; Bovino,~S.; Caselli,~P.; Bovolenta,~G.; Vogt-Geisse,~S.;
  Ercolano,~B. A novel framework for studying the impact of binding energy
  distributions on the chemistry of dust grains. \emph{Astronomy \&
  Astrophysics} \textbf{2020}, \emph{643}, A155, Publisher: EDP Sciences\relax
\mciteBstWouldAddEndPuncttrue
\mciteSetBstMidEndSepPunct{\mcitedefaultmidpunct}
{\mcitedefaultendpunct}{\mcitedefaultseppunct}\relax
\EndOfBibitem
\bibitem[Bovolenta \latin{et~al.}(2020)Bovolenta, Bovino, VÃ¶hringer-Martinez,
  Saez, Grassi, and Vogt-Geisse]{bovolenta_high_2020}
Bovolenta,~G.; Bovino,~S.; VÃ¶hringer-Martinez,~E.; Saez,~D.~A.; Grassi,~T.;
  Vogt-Geisse,~S. High level ab initio binding energy distribution of molecules
  on interstellar ices: {Hydrogen} fluoride. \emph{Molecular Astrophysics}
  \textbf{2020}, 100095\relax
\mciteBstWouldAddEndPuncttrue
\mciteSetBstMidEndSepPunct{\mcitedefaultmidpunct}
{\mcitedefaultendpunct}{\mcitedefaultseppunct}\relax
\EndOfBibitem
\bibitem[Galano(2007)]{galano_mechanism_2007}
Galano,~A. Mechanism of {OH} {Radical} {Reactions} with {HCN} and {CH3CN}:â
  {OH} {Regeneration} in the {Presence} of {O2}. \emph{The Journal of Physical
  Chemistry A} \textbf{2007}, \emph{111}, 5086--5091, Publisher: American
  Chemical Society\relax
\mciteBstWouldAddEndPuncttrue
\mciteSetBstMidEndSepPunct{\mcitedefaultmidpunct}
{\mcitedefaultendpunct}{\mcitedefaultseppunct}\relax
\EndOfBibitem
\bibitem[Bunkan \latin{et~al.}(2013)Bunkan, Liang, Pilling, and
  Nielsen]{bunkan_theoretical_2013}
Bunkan,~A. J.~C.; Liang,~C.-H.; Pilling,~M.~J.; Nielsen,~C.~J. Theoretical and
  experimental study of the {OH} radical reaction with {HCN}. \emph{Molecular
  Physics} \textbf{2013}, \emph{111}, 1589--1598, Publisher: Taylor \& Francis
  \_eprint: https://doi.org/10.1080/00268976.2013.802036\relax
\mciteBstWouldAddEndPuncttrue
\mciteSetBstMidEndSepPunct{\mcitedefaultmidpunct}
{\mcitedefaultendpunct}{\mcitedefaultseppunct}\relax
\EndOfBibitem
\bibitem[Garrod(2013)]{garrod_three-phase_2013}
Garrod,~R.~T. A {THREE}-{PHASE} {CHEMICAL} {MODEL} {OF} {HOT} {CORES}: {THE}
  {FORMATION} {OF} {GLYCINE}. \emph{The Astrophysical Journal} \textbf{2013},
  \emph{765}, 60, Publisher: IOP Publishing\relax
\mciteBstWouldAddEndPuncttrue
\mciteSetBstMidEndSepPunct{\mcitedefaultmidpunct}
{\mcitedefaultendpunct}{\mcitedefaultseppunct}\relax
\EndOfBibitem
\bibitem[Tsuge and Watanabe(2021)Tsuge, and Watanabe]{tsuge_behavior_2021}
Tsuge,~M.; Watanabe,~N. Behavior of {Hydroxyl} {Radicals} on {Water} {Ice} at
  {Low} {Temperatures}. \emph{Accounts of Chemical Research} \textbf{2021},
  \emph{54}, 471--480, Publisher: American Chemical Society\relax
\mciteBstWouldAddEndPuncttrue
\mciteSetBstMidEndSepPunct{\mcitedefaultmidpunct}
{\mcitedefaultendpunct}{\mcitedefaultseppunct}\relax
\EndOfBibitem
\bibitem[Miyazaki \latin{et~al.}(2022)Miyazaki, Tsuge, Hidaka, Nakai, and
  Watanabe]{miyazaki_direct_2022}
Miyazaki,~A.; Tsuge,~M.; Hidaka,~H.; Nakai,~Y.; Watanabe,~N. Direct
  {Determination} of the {Activation} {Energy} for {Diffusion} of {OH}
  {Radicals} on {Water} {Ice}. \emph{The Astrophysical Journal Letters}
  \textbf{2022}, \emph{940}, L2\relax
\mciteBstWouldAddEndPuncttrue
\mciteSetBstMidEndSepPunct{\mcitedefaultmidpunct}
{\mcitedefaultendpunct}{\mcitedefaultseppunct}\relax
\EndOfBibitem
\bibitem[Bovolenta \latin{et~al.}(2022)Bovolenta, Vogt-Geisse, Bovino, and
  Grassi]{bovolenta_binding_2022}
Bovolenta,~G.~M.; Vogt-Geisse,~S.; Bovino,~S.; Grassi,~T. Binding {Energy}
  {Evaluation} {Platform}: {A} {Database} of {Quantum} {Chemical} {Binding}
  {Energy} {Distributions} for the {Astrochemical} {Community}. \emph{The
  Astrophysical Journal Supplement Series} \textbf{2022}, \emph{262}, 17,
  Publisher: The American Astronomical Society\relax
\mciteBstWouldAddEndPuncttrue
\mciteSetBstMidEndSepPunct{\mcitedefaultmidpunct}
{\mcitedefaultendpunct}{\mcitedefaultseppunct}\relax
\EndOfBibitem
\bibitem[Shingledecker \latin{et~al.}(2024)Shingledecker, Vogt-Geisse, Mifsud,
  and Ioppolo]{shingledecker_chapter_2024}
Shingledecker,~C.~N.; Vogt-Geisse,~S.; Mifsud,~D.~V.; Ioppolo,~S. In
  \emph{Astrochemical {Modeling}}; Bovino,~S., Grassi,~T., Eds.; Elsevier,
  2024; pp 71--115\relax
\mciteBstWouldAddEndPuncttrue
\mciteSetBstMidEndSepPunct{\mcitedefaultmidpunct}
{\mcitedefaultendpunct}{\mcitedefaultseppunct}\relax
\EndOfBibitem
\bibitem[Sure and Grimme(2013)Sure, and Grimme]{sure_corrected_2013}
Sure,~R.; Grimme,~S. Corrected small basis set {Hartree}-{Fock} method for
  large systems. \emph{Journal of Computational Chemistry} \textbf{2013},
  \emph{34}, 1672--1685, \_eprint:
  https://onlinelibrary.wiley.com/doi/pdf/10.1002/jcc.23317\relax
\mciteBstWouldAddEndPuncttrue
\mciteSetBstMidEndSepPunct{\mcitedefaultmidpunct}
{\mcitedefaultendpunct}{\mcitedefaultseppunct}\relax
\EndOfBibitem
\bibitem[Zhao and Truhlar(2004)Zhao, and Truhlar]{zhao_hybrid_2004}
Zhao,~Y.; Truhlar,~D.~G. Hybrid {Meta} {Density} {Functional} {Theory}
  {Methods} for {Thermochemistry}, {Thermochemical} {Kinetics}, and
  {Noncovalent} {Interactions}: {The} {MPW1B95} and {MPWB1K} {Models} and
  {Comparative} {Assessments} for {Hydrogen} {Bonding} and van der {Waals}
  {Interactions}. \emph{The Journal of Physical Chemistry A} \textbf{2004},
  \emph{108}, 6908--6918\relax
\mciteBstWouldAddEndPuncttrue
\mciteSetBstMidEndSepPunct{\mcitedefaultmidpunct}
{\mcitedefaultendpunct}{\mcitedefaultseppunct}\relax
\EndOfBibitem
\bibitem[Weigend and Ahlrichs(2005)Weigend, and
  Ahlrichs]{weigend_balanced_2005}
Weigend,~F.; Ahlrichs,~R. Balanced basis sets of split valence, triple zeta
  valence and quadruple zeta valence quality for {H} to {Rn}: {Design} and
  assessment of accuracy. \emph{Physical Chemistry Chemical Physics}
  \textbf{2005}, \emph{7}, 3297\relax
\mciteBstWouldAddEndPuncttrue
\mciteSetBstMidEndSepPunct{\mcitedefaultmidpunct}
{\mcitedefaultendpunct}{\mcitedefaultseppunct}\relax
\EndOfBibitem
\bibitem[Grimme \latin{et~al.}(2011)Grimme, Ehrlich, and
  Goerigk]{grimme_effect_2011}
Grimme,~S.; Ehrlich,~S.; Goerigk,~L. Effect of the damping function in
  dispersion corrected density functional theory. \emph{Journal of
  Computational Chemistry} \textbf{2011}, \emph{32}, 1456--1465\relax
\mciteBstWouldAddEndPuncttrue
\mciteSetBstMidEndSepPunct{\mcitedefaultmidpunct}
{\mcitedefaultendpunct}{\mcitedefaultseppunct}\relax
\EndOfBibitem
\bibitem[Seritan \latin{et~al.}(2021)Seritan, Bannwarth, Fales, Hohenstein,
  Isborn, Kokkila-Schumacher, Li, Liu, Luehr, Snyder~Jr., Song, Titov,
  Ufimtsev, Wang, and MartÃ­nez]{seritan_terachem_2021}
Seritan,~S.; Bannwarth,~C.; Fales,~B.~S.; Hohenstein,~E.~G.; Isborn,~C.~M.;
  Kokkila-Schumacher,~S. I.~L.; Li,~X.; Liu,~F.; Luehr,~N.; Snyder~Jr.,~J.~W.;
  Song,~C.; Titov,~A.~V.; Ufimtsev,~I.~S.; Wang,~L.-P.; MartÃ­nez,~T.~J.
  {TeraChem}: {A} graphical processing unit-accelerated electronic structure
  package for large-scale ab initio molecular dynamics. \emph{WIREs
  Computational Molecular Science} \textbf{2021}, \emph{11}, e1494, \_eprint:
  https://onlinelibrary.wiley.com/doi/pdf/10.1002/wcms.1494\relax
\mciteBstWouldAddEndPuncttrue
\mciteSetBstMidEndSepPunct{\mcitedefaultmidpunct}
{\mcitedefaultendpunct}{\mcitedefaultseppunct}\relax
\EndOfBibitem
\bibitem[Turney \latin{et~al.}(2012)Turney, Simmonett, Parrish, Hohenstein,
  Evangelista, Fermann, Mintz, Burns, Wilke, Abrams, Russ, Leininger, Janssen,
  Seidl, Allen, Schaefer, King, Valeev, Sherrill, and
  Crawford]{turney_psi4_2012}
Turney,~J.~M. \latin{et~al.}  Psi4: an open-source ab initio electronic
  structure program. \emph{WIREs Computational Molecular Science}
  \textbf{2012}, \emph{2}, 556--565, \_eprint:
  https://onlinelibrary.wiley.com/doi/pdf/10.1002/wcms.93\relax
\mciteBstWouldAddEndPuncttrue
\mciteSetBstMidEndSepPunct{\mcitedefaultmidpunct}
{\mcitedefaultendpunct}{\mcitedefaultseppunct}\relax
\EndOfBibitem
\bibitem[Smith \latin{et~al.}(2021)Smith, Altarawy, Burns, Welborn, Naden,
  Ward, Ellis, Pritchard, and Crawford]{smith_molssi_2021}
Smith,~D. G.~A.; Altarawy,~D.; Burns,~L.~A.; Welborn,~M.; Naden,~L.~N.;
  Ward,~L.; Ellis,~S.; Pritchard,~B.~P.; Crawford,~T.~D. The {MolSSI}
  {QCArchive} project: {An} open-source platform to compute, organize, and
  share quantum chemistry data. \emph{WIREs Computational Molecular Science}
  \textbf{2021}, \emph{11}, e1491, \_eprint:
  https://onlinelibrary.wiley.com/doi/pdf/10.1002/wcms.1491\relax
\mciteBstWouldAddEndPuncttrue
\mciteSetBstMidEndSepPunct{\mcitedefaultmidpunct}
{\mcitedefaultendpunct}{\mcitedefaultseppunct}\relax
\EndOfBibitem
\bibitem[Wang and Song(2016)Wang, and Song]{wang_geometry_2016}
Wang,~L.-P.; Song,~C. Geometry optimization made simple with translation and
  rotation coordinates. \emph{The Journal of Chemical Physics} \textbf{2016},
  \emph{144}, 214108\relax
\mciteBstWouldAddEndPuncttrue
\mciteSetBstMidEndSepPunct{\mcitedefaultmidpunct}
{\mcitedefaultendpunct}{\mcitedefaultseppunct}\relax
\EndOfBibitem
\bibitem[Becke(1993)]{becke_new_1993}
Becke,~A.~D. A new mixing of {Hartree}â{Fock} and local density-functional
  theories. \emph{The Journal of Chemical Physics} \textbf{1993}, \emph{98},
  1372--1377\relax
\mciteBstWouldAddEndPuncttrue
\mciteSetBstMidEndSepPunct{\mcitedefaultmidpunct}
{\mcitedefaultendpunct}{\mcitedefaultseppunct}\relax
\EndOfBibitem
\bibitem[Boese and Martin(2004)Boese, and Martin]{boese_development_2004}
Boese,~A.~D.; Martin,~J. M.~L. Development of density functionals for
  thermochemical kinetics. \emph{The Journal of Chemical Physics}
  \textbf{2004}, \emph{121}, 3405--3416\relax
\mciteBstWouldAddEndPuncttrue
\mciteSetBstMidEndSepPunct{\mcitedefaultmidpunct}
{\mcitedefaultendpunct}{\mcitedefaultseppunct}\relax
\EndOfBibitem
\bibitem[Warden \latin{et~al.}(2020)Warden, Smith, Burns, Bozkaya, and
  Sherrill]{warden_efficient_2020}
Warden,~C.~E.; Smith,~D. G.~A.; Burns,~L.~A.; Bozkaya,~U.; Sherrill,~C.~D.
  Efficient and automated computation of accurate molecular geometries using
  focal-point approximations to large-basis coupled-cluster theory. \emph{The
  Journal of Chemical Physics} \textbf{2020}, \emph{152}, 124109, Publisher:
  American Institute of Physics\relax
\mciteBstWouldAddEndPuncttrue
\mciteSetBstMidEndSepPunct{\mcitedefaultmidpunct}
{\mcitedefaultendpunct}{\mcitedefaultseppunct}\relax
\EndOfBibitem
\bibitem[Woon(2002)]{woon_ab_2002}
Woon,~D.~E. Ab initio quantum chemical studies of reactions in astrophysical
  ices. 4. {Reactions} in ices involving {HCOOH}, {CH2NH}, {HCN}, {HNC}, {NH3},
  and {H2O}. \emph{International Journal of Quantum Chemistry} \textbf{2002},
  \emph{88}, 226--235\relax
\mciteBstWouldAddEndPuncttrue
\mciteSetBstMidEndSepPunct{\mcitedefaultmidpunct}
{\mcitedefaultendpunct}{\mcitedefaultseppunct}\relax
\EndOfBibitem
\bibitem[Baiano \latin{et~al.}(2022)Baiano, Lupi, Barone, and
  Tasinato]{baiano_gliding_2022}
Baiano,~C.; Lupi,~J.; Barone,~V.; Tasinato,~N. Gliding on {Ice} in {Search} of
  {Accurate} and {Cost}-{Effective} {Computational} {Methods} for
  {Astrochemistry} on {Grains}: {The} {Puzzling} {Case} of the {HCN}
  {Isomerization}. \emph{Journal of Chemical Theory and Computation}
  \textbf{2022}, \emph{18}, 3111--3121, Publisher: American Chemical
  Society\relax
\mciteBstWouldAddEndPuncttrue
\mciteSetBstMidEndSepPunct{\mcitedefaultmidpunct}
{\mcitedefaultendpunct}{\mcitedefaultseppunct}\relax
\EndOfBibitem
\bibitem[Tinacci \latin{et~al.}(2023)Tinacci, Germain, Pantaleone, Ceccarelli,
  Balucani, and Ugliengo]{Tinacci_2023}
Tinacci,~L.; Germain,~A.; Pantaleone,~S.; Ceccarelli,~C.; Balucani,~N.;
  Ugliengo,~P. Theoretical Water Binding Energy Distribution and Snowline in
  Protoplanetary Disks. \emph{The Astrophysical Journal} \textbf{2023},
  \emph{951}, 32\relax
\mciteBstWouldAddEndPuncttrue
\mciteSetBstMidEndSepPunct{\mcitedefaultmidpunct}
{\mcitedefaultendpunct}{\mcitedefaultseppunct}\relax
\EndOfBibitem
\end{mcitethebibliography}
\end{document}